\documentclass[10pt,journal,compsoc]{IEEEtran}
\usepackage{flushend}
\ifCLASSOPTIONcompsoc
  \usepackage[nocompress]{cite}
\else
  \usepackage{cite}
\fi

\ifCLASSINFOpdf
   \usepackage[pdftex]{graphicx}
\else
\fi

\ifCLASSINFOpdf
\else
\fi

\usepackage{multirow}
\usepackage{makecell}
\usepackage[cmintegrals]{newtxmath}
\usepackage{bm}
\usepackage{algorithm}
\usepackage{algorithmic}
\usepackage{indentfirst}
  
\usepackage{amsmath,amssymb}
\usepackage{array}
\usepackage{pifont}
\usepackage{mathrsfs}
\usepackage{cite}
\usepackage{tikz}
\usepackage{color}
\usepackage{amsmath}
\usepackage{colortbl}

\newcommand*{\circled}[1]{\lower.7ex\hbox{\tikz\draw (0pt, 0pt)%
    circle (.5em) node {\makebox[1em][c]{\small #1}};}}

\ifCLASSOPTIONcompsoc
 \usepackage[caption=false,font=normalsize,labelfont=sf,textfont=sf]{subfig}
\else
 \usepackage[caption=false,font=footnotesize]{subfig}
\fi

\begin{document}
\title{M$^{3}$FAS: An Accurate and Robust MultiModal Mobile Face Anti-Spoofing System}

\author{Chenqi~Kong, Kexin~Zheng, Yibing~Liu,
Shiqi~Wang,~\IEEEmembership{Senior Member,~IEEE},\\
Anderson~Rocha,~\IEEEmembership{Fellow,~IEEE},
and Haoliang~Li,~\IEEEmembership{Member,~IEEE}
\IEEEcompsocitemizethanks{\IEEEcompsocthanksitem C. Kong, Y. Liu, and S. Wang are with the Department
of Computer Science, City University of Hong Kong, Hong Kong, China.\protect\\
E-mail: (cqkong2-c@my.cityu.edu.hk; lyibing112@gmail.com; shiqwang@cityu.edu.hk)
\IEEEcompsocthanksitem K. Zheng is with the Department of Computer Science and Engineering, Hong Kong University of Science and Technology, Hong Kong, China.\protect\\
E-mail: kzhengaj@connect.ust.hk
\IEEEcompsocthanksitem A. Rocha is with the Artificial Intelligence Lab. (\texttt{Recod.ai}) at the University of Campinas, Campinas 13084-851, Brazil.\protect\\
E-mail: anderson.rocha@ic.unicamp.br
\IEEEcompsocthanksitem H. Li is with the Department
of Electrical and Engineering, City University of Hong Kong, Hong Kong, China. H. Li is the corresponding author.\protect\\
E-mail: haoliang.li@cityu.edu.hk}}

\markboth{Submitted to IEEE TRANSACTIONS ON DEPENDABLE AND SECURE COMPUTING}%
{Shell \MakeLowercase{\textit{et al.}}: Bare Demo of IEEEtran.cls for Computer Society Journals}

\IEEEtitleabstractindextext{%
\begin{abstract}
Face presentation attacks (FPA), also known as face spoofing, have brought increasing concerns to the public through various malicious applications, such as financial fraud and privacy leakage. Therefore, safeguarding face recognition systems against FPA is of utmost importance. Although existing learning-based face anti-spoofing (FAS) models can achieve outstanding detection performance, they lack generalization capability and suffer significant performance drops in unforeseen environments. Many methodologies seek to use auxiliary modality data ($e.g.$, depth and infrared maps) during the presentation attack detection (PAD) to address this limitation. However, these methods can be limited since (1) they require specific sensors such as depth and infrared cameras for data capture, which are rarely available on commodity mobile devices, and (2) they cannot work properly in practical scenarios when either modality is missing or of poor quality. In this paper, we devise an accurate and robust \textbf{M}ulti\textbf{M}odal \textbf{M}obile \textbf{F}ace \textbf{A}nti-\textbf{S}poofing system named \textbf{M$^{3}$FAS} to overcome the issues above.
The primary innovation of this work lies in the following aspects: (1) To achieve robust PAD, our system combines visual and auditory modalities using three commonly available sensors: camera, speaker, and microphone; (2) We design a novel two-branch neural network with three hierarchical feature aggregation modules to perform cross-modal feature fusion; (3). We propose a multi-head training strategy, allowing the model to output predictions from the vision, acoustic, and fusion heads, resulting in a more flexible PAD. Extensive experiments have demonstrated the accuracy, robustness, and flexibility of M$^{3}$FAS under various challenging experimental settings. \textcolor{black}{The source code and dataset are available at: https://github.com/ChenqiKONG/M3FAS/}.



\end{abstract}

}

\maketitle
\IEEEdisplaynontitleabstractindextext
\IEEEpeerreviewmaketitle

\IEEEraisesectionheading{\section{Introduction}\label{sec:introduction}}

\IEEEPARstart{A}{utomatic} face recognition (AFR) systems have been prevalently deployed on mobile devices and play a vital role around the globe. It is reported that the market of AFR will reach USD 3.35B by 2024~\cite{AFR_market}. Despite AFR's extraordinary success, face presentation attacks (FPA), also known as face spoofing, have recently posed high-security risks over impersonation, financial fraud, and privacy leakage. 2D FPA, comprising photo print and video replay attacks, is the most detrimental and notorious attack type due to its low costs~\cite{patel2016secure}. Malicious attackers can easily launch it by accessing the target person's face images/videos on social media and presenting them to the target AFR systems. The abuse of 2D FPA will certainly lead to trust destruction and tangible concerns in the long run. Therefore, safeguarding AFR systems from 2D FPA and suppressing the pressing security concerns is of utmost importance. 

\begin{figure}[ht]
\centering
\includegraphics[scale=0.255]{ 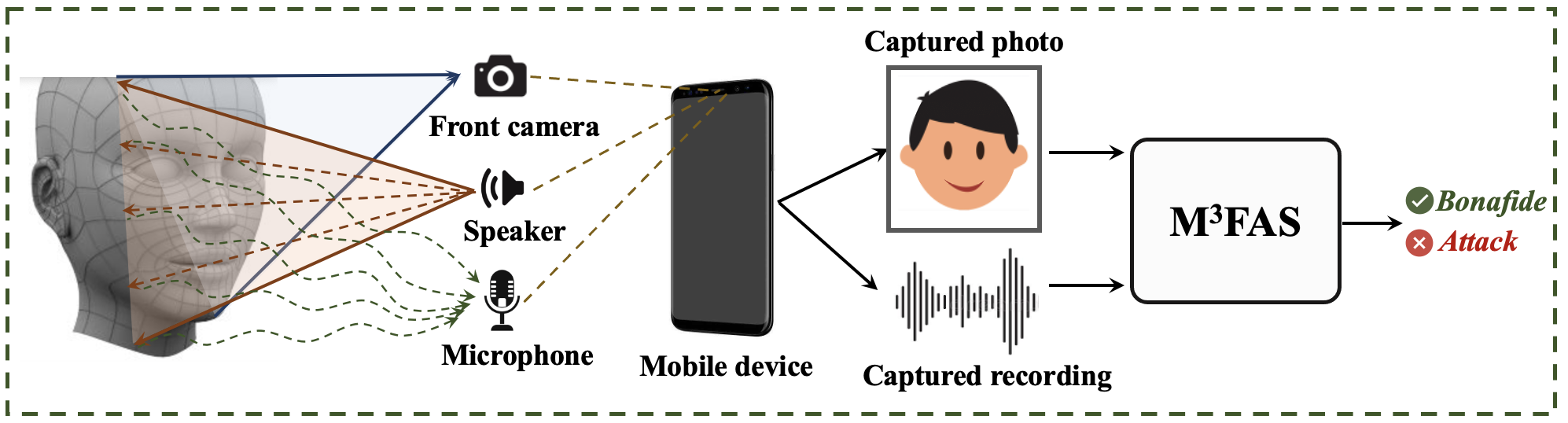}
\caption{Illustration of M$^3$FAS system. The mobile device employs the front camera to capture the input RGB face photo. Meanwhile, the top speaker emits a customized acoustic signal, and the microphone collects the reflected signal modulated by the live/spoof face. The captured face picture and acoustic recording are sequentially fed forward to M$^3$FAS for final decision-making.}
\label{teaser}
\end{figure}

The past decade has witnessed important progress in presentation attack detection (PAD) methodologies, which can generally be classified into two categories: traditional handcrafted and deep learning-based methods. On the one hand, the former aims at extracting handcrafted features from RGB face images, such as LBP~\cite{maatta2011face,de2012lbp}, HoG~\cite{komulainen2013context}, IDA~\cite{wen2015face}, LPQ~\cite{boulkenafet2016face}, SIFT~\cite{patel2016secure}, SURF~\cite{boulkenafet2016face}, among others. Although computationally efficient, they tend to suffer from limited PAD accuracy. On the other hand, many learning-based PAD models have been proposed in the era of artificial intelligence. These data-driven methods can achieve outstanding face liveness detection performance while lacking generalization capability to the data captured from unseen environments. To bridge such domain gaps, recent methods~\cite{liu2021face, sun2020face, wang2020deep, pinto2020leveraging, liu2021data, nikisins2019domain} propose to perform PAD by simultaneously using RGB images and auxiliary modality data ($e.g.$, depth maps, infrared maps). These auxiliary modalities contain rich inherent spoof cues, thus can effectively mitigate the overfitting problem and improve the robustness.
Nonetheless, existing multimodal FAS methodologies still pose two drawbacks: (1) they require specific sensors ($e.g.$, depth camera, infrared camera) for auxiliary modality data acquisition, which are rarely available on commercial-off-the-shelf (COTS) devices, especially on mobile devices; 
(2) they may not work properly when one modality is missing or of poor quality in some practical scenarios, such as hardware failure and poor environmental conditions.

In this paper, we devise an accurate and robust multimodal mobile face anti-spoofing system, referred to as \textbf{M$^{3}$FAS}, to address the issues above properly. 
M$^{3}$FAS assembles RGB data and acoustic signals and designs three hierarchical feature aggregation modules to perform secure and robust PAD. 
We leverage three binary cross-entropy losses to jointly supervise the training of the whole framework. During the training phase, M$^{3}$FAS generates predictions from three separate heads: the RGB head, the acoustic head, and the fusion head. This enables the system to adapt flexibly in the inference stage, even in diverse and demanding practical environments, where one of the modalities could be missing or of poor quality.
Fig.~\ref{teaser} depicts the proposed multimodal PAD progress. First, the front camera captures the face picture. Meanwhile, the top speaker emits a custom-tailored high-frequency signal to visit the input face, and the top microphone will record the reflected acoustic signal. Then, the M$^{3}$FAS system jointly combines visual and acoustic features of the input query to perform PAD. In this vein, the properties of our system are summarized as follows:
\begin{itemize}
    \item \textbf{Dependable}: Prior arts have shown that acoustic signal is reliable in extracting representative dynamic or static biometric patterns. It has been demonstrated effective in diverse biometric applications, such as finger tracking~\cite{fingerIO2016}, gesture recognition~\cite{ultragesture2018, liao2021smart, yu2019rfid}, and face authentication~\cite{xu2021rface, zhou2018echoprint}. Encouraged by the demonstrated success, this work combines visual and auditory modalities, which are dependable for spoof feature extraction. 
    \item \textbf{Applicable}: The ubiquitous availability of a front camera, speaker, and microphone on COTS mobile devices enables the easy deployment of M$^{3}$FAS in real-world applications. Besides, the model equipped with auditory modality is more computationally-efficient than other auxiliary modalities ($e.g.$, depth map) due to its one-dimension scale nature, thus contributing to fast response in practical applications. 
    \item \textbf{Robust}: RGB-based FAS model is fragile against attacks launched in unseen environments. The captured acoustic signal reflects much geometric facial information of the query, which has been largely ignored in the RGB modality.  
    As such, M$^{3}$FAS complementarily combines visual texture features and acoustic geometric facial features, thereby achieving a more robust PAD performance. 
    \item \textbf{Flexible}: As illustrated in Fig.~\ref{framework}, three separate heads are designed to output three predictions in the training phase. Thus, M$^{3}$FAS can flexibly perform PAD even if one modality is missing in some extreme scenarios. 
\end{itemize}


We first employ four smartphones to collect a large-scale multimodal face spoofing database from 30 participants with diverse variables.
Then, grounded on the database, we conduct extensive experiments under both intra- and cross-domain settings to verify the effectiveness and robustness of the proposed M$^{3}$FAS system. \textcolor{black}{Compared with our previous work~\cite{kong2022beyond}, the primary novelties and differences of this work are summarized as follows: \textbf{(1) Learning scheme:} Unlike our previous work~\cite{kong2022beyond}, which mainly focuses on unimodal FAS, M$^{3}$FAS conducts a comprehensive study into the audio-visual multimodal FAS problem. In this work, we demonstrated the effectiveness of the proposed multimodal fusion scheme through extensive experiments, encompassing cross-subject, cross-device, and cross-environment variable evaluations (distance, noise, head pose), as well as assessments under various image distortions; \textbf{(2) Methodology:} Our designed FAS framework differs from previous work~\cite{kong2022beyond} in two key aspects: (a) Hierarchical cross-attention module (HCAM): HCAM incorporates two cross-modality attention modules to align the features from visual and auditory modalities. Additionally, HCAM introduces a hierarchical feature accumulation strategy, combining multimodal features across low, middle, and high levels. This enables the hierarchical interaction and fusion of features from different levels, thereby enhancing the final FAS performance;
(b) Multi-head learning scheme: We propose a multi-head learning strategy for the visual-auditory multimodal FAS task. This design enables a more flexible FAS during the inference stage. Furthermore, experimental results demonstrate the effectiveness of the proposed multi-head learning scheme in mitigating overfitting and improving PAD performance; \textbf{(3) Database:} In contrast to our previous work~\cite{kong2022beyond}, which introduced an acoustic-based FAS database, this work takes a significant step forward by establishing a multimodal FAS database named Echoface-Spoof. This large-scale and diverse database incorporates visual and auditory modalities and contains approximately 250,000 pairs of face images and acoustic signal data with various variables, including device, distance, ambient noise, pitch, and image distortion. Echoface-Spoof can facilitate the development of more generalized and robust multimodal FAS frameworks.}

The main contributions of this work are:
\begin{itemize}
    \item[$\bullet$] We introduce a large-scale high-diversity multimodal FAS database (Echoface-Spoof) using four smartphones, which is significantly extended from our previous work \cite{kong2022beyond} by incorporating different modalities and more 
    environmental variables. It includes around 250,000 face image and acoustic signal data pairs with diverse variables over capture device, distance, ambient noise, pitch, and image distortion.
    \item[$\bullet$] We present a novel multimodal FAS framework that complementarily harnesses acoustic and visual features via hierarchical cross-attention modules. We design three separate heads for the vision, acoustic, and combine channels to conduct binary classifications in the training stage, thereby enabling a flexible FAS in the inference phase. The proposed joint training can effectively boost detection performance compared with the different training strategies.
    \item[$\bullet$] 
    The system can achieve accurate and robust PAD performance under intra- and cross-domain settings. Extensive experimental results have demonstrated the accuracy, robustness, and flexibility of our M$^3$FAS system. 
\end{itemize}

Section 2 reviews related work on FAS and acoustic-based biometric applications. Section 3 presents the signal configuration and dataset acquisition process. Section 4  details the proposed M$^3$FAS system. Section 5 reports comprehensive evaluation results under diverse experimental settings. Finally, Section 6 concludes this paper and discusses current limitations and possible future research directions.

\begin{figure*}[ht]
\centering
\includegraphics[scale=0.53]{ 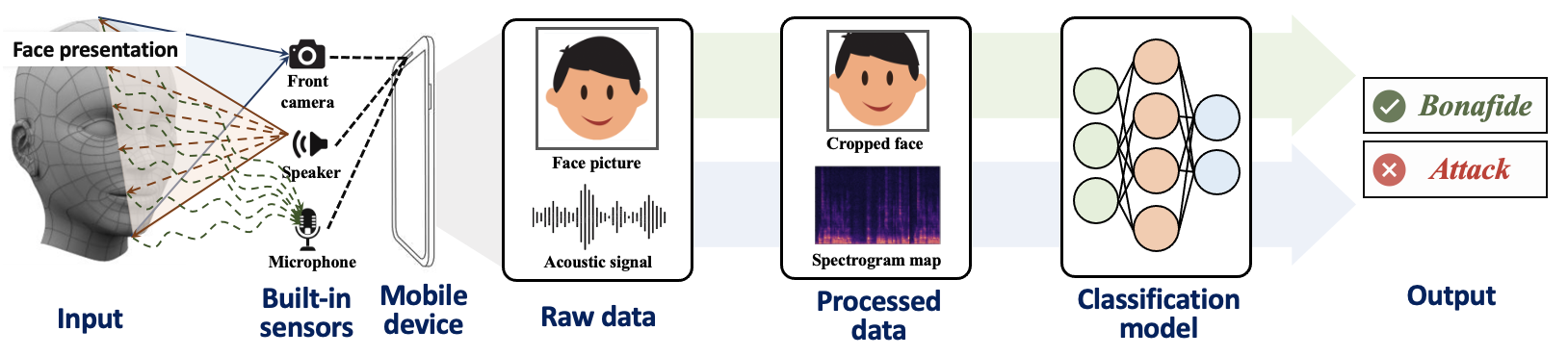}
\vspace{-0.2cm}
\caption{Pipeline of M$^3$FAS system. The mobile device employs built-in sensors to capture the input data pair: a face picture and a raw acoustic signal segment. 
These inputs are then preprocessed by face and signal extraction algorithms to obtain the cropped face image and spectrogram map. Finally, the classification model produces the final decision.} 
\label{pipeline}
\end{figure*}

\section{Related Work}
In this section, we broadly review existing works on presentation attack detection (PAD), including handcrafted, learning-based, multi-modality PAD methodologies, as well as PAD systems and their applications. 

\subsection{Handcrafted PAD methodologies} Presentation attack detection methodologies can be classified into two categories: handcrafted and learning-based. Generally speaking, the print, replay, and recapture process will inevitably introduce certain image distortions, such as moire pattern, printed noise, and color mismatch ~\cite{kong2022digital}. In this vein, traditional handcrafted methods exploit a wide variety of handcrafted feature extractors ($e.g.$, LBP~\cite{maatta2011face,de2012lbp}, LPQ~\cite{boulkenafet2016face}, HoG~\cite{komulainen2013context}, SIFT~\cite{patel2016secure}, SURF~\cite{boulkenafet2016face}, IDA~\cite{wen2015face}, and DoG~\cite{tan2010face}) to expose the inherent artifacts of face spoofing. For instance, LBP~\cite{maatta2011face,de2012lbp} is a local texture descriptor that captures 
texture artifacts from RGB images to discriminate the bonafide from attacks. SIFT~\cite{patel2016secure} has been widely taken as image representations in various vision tasks. Thanks to its robustness against image rotation, scale, and translation,  \cite{patel2016secure} demonstrated that SIFT was also effective in face liveness detection. DoG~\cite{tan2010face} and HoG~\cite{komulainen2013context} filters are edge and gradient-based descriptors that highlight the high-frequency information in images, hence powering the PAD performance. \textbf{\textit{Typically, handcrafted models are computation-effective but suffer from low detection accuracy and limited generalization capability.}}

\vspace{-0.25cm}
\subsection{Learning-based PAD methodologies} Thanks to the advent of artificial intelligence and deep learning, past decades have witnessed substantial efforts in developing learning-based PAD models. For the first time, Yang $et~al.$ \cite{yang2014learn} employed the convolutional neural network (CNN) to conduct PAD and achieved good detection accuracy. Lucena $et~al.$ demonstrated that VGG16 \cite{simonyan2014very} pretrained on ImageNet~\cite{russakovsky2015imagenet} could effectively save the training time and computational resources on the task of face liveness detection. For video replay attack detection, long short-term memory (LSTM)~\cite{ge2020face, xu2015learning} and recurrent neural network (RNN)~\cite{muhammad2019face} architectures have been incorporated to capture the artifacts in terms of temporal inconsistency. Follow-up works such as \cite{yang2019face} designed a spatiotemporal face anti-spoofing network to smartly fuse spatial and temporal spoofing cues, achieving an accurate PAD performance. Powered by powerful deep learning tools, learning-based methodologies have been increasingly accurate on PAD. Yu $et~al.$ \cite{yu2020fas} firstly propose a neural architecture search-based face anti-spoofing method and achieves SOTA performance on nine datasets. \textbf{\textit{Nonetheless, these data-driven methods are prone to overfitting the training data, resulting in limited generalization capability to unforeseen domains.}}    

\vspace{-0.25cm}
\subsection{Multi-modality PAD methodologies}
Binary supervision can cause severe overfitting problems. For robustness, recent methods proposed to conduct presentation attack detection via multi-modality fusion, which captures informative spoofing cues from different modalities and fuses them in a complementary fashion. Depth maps reflect much 3D geometric facial information. Thus they have been widely used in 2D PAD and achieved better robustness \cite{atoum2017face, liu2018learning, wang2020deep, yu2020searching, cai2022learning, nikisins2019domain, liu2021face, cai2023s, shen2019facebagnet, yu2020multi, liu2021casia, luo2023beyond, kong2023enhancing, wan2020multi, liu2021cross, liu2023ma}. Remote photoplethysmography (rPPG) signals can effectively expose heart rhythm patterns and accurately discriminate bonafide from attacks \cite{li2016generalized, yu2022benchmarking, lin2019face, liu2018learning, yu2019remote, liu2019multi}. Besides, researchers \cite{kuang2019multi, liu2021data, nikisins2019domain, liu2021face, jiang2020face, shen2019facebagnet, yu2020multi, zhang2019feathernets} recently demonstrated that near-infrared (NIR) information exposed more inherent spoofing clues since it measures the heat radiation amount of a live face. Some works also seek to use binary masks \cite{liu2020disentangling, hossain2020deeppixbis, yu2020auto, george2019deep, sun2020face, liu2019deep} and auxiliary geometric information \cite{kim2019basn, zhang2020celeba, yu2020face} to boost the generalization capability. \textbf{\textit{However, most of these methods require extra hardware that is rarely available on mobile devices, causing difficulties when deployed.}}

\vspace{-0.1cm}
\subsection{PAD systems and mobile applications}
\vspace{-0.1cm}
The availability of various sensors built into COTS mobile devices enables multimodal presentation attack detection in practical scenarios. In the past five years, many mobile-oriented presentation attack detection systems have been devised, among which FaceHeart~\cite{chen2017your} simultaneously takes a face video with the front camera and a fingertip video with the rear camera. Then, it compares the rPPG similarity between these two videos to conduct the decision-making. Apple's highly touted FaceID~\cite{apple} has demonstrated extraordinary accuracy for both face authentication and face presentation attack detection. However, it requires expensive infrared dot projectors
and dedicated cameras and does not align well with the smartphone industry’s desire to maximize screen space. To solve this, FaceRevelio~\cite{farrukh2020facerevelio} uses the screen to illuminate the face from different directions and captures the face pictures with a front camera. Face liveness detection can be subsequently achieved by reconstructing the 3D face structure. Face Flashing~\cite{tang2018face} detected the face liveness by randomly flashing well-designed pictures on a screen and analyzing the reflected light. However, it requires the user's expressions. On the other hand, recent success on acoustic signal has exploded into a plethora of practical PAD applications \cite{zhou2021securing, zhou2018echoprint, chen2019echoface, kong2022beyond}. EchoPrint~\cite{zhou2018echoprint}, for the first time, combined visual and acoustic signals to conduct face authentication while leaving face liveness detection as a pressing issue. Follow-up works, including EchoFace~\cite{chen2019echoface}, RFace~\cite{xu2021rface}, and EchoFAS~\cite{kong2022beyond}, mainly focus on the acoustic-based PAD. \textbf{\textit{However, the multi-modality PAD has been largely understudied.}} 

In this paper, we devise a face anti-spoofing system M$^{3}$FAS, incorporating visual and acoustic spoofing features to conduct a more secure, robust, and flexible PAD. 



\begin{figure*}[ht]
\centering
\includegraphics[scale=0.38]{ 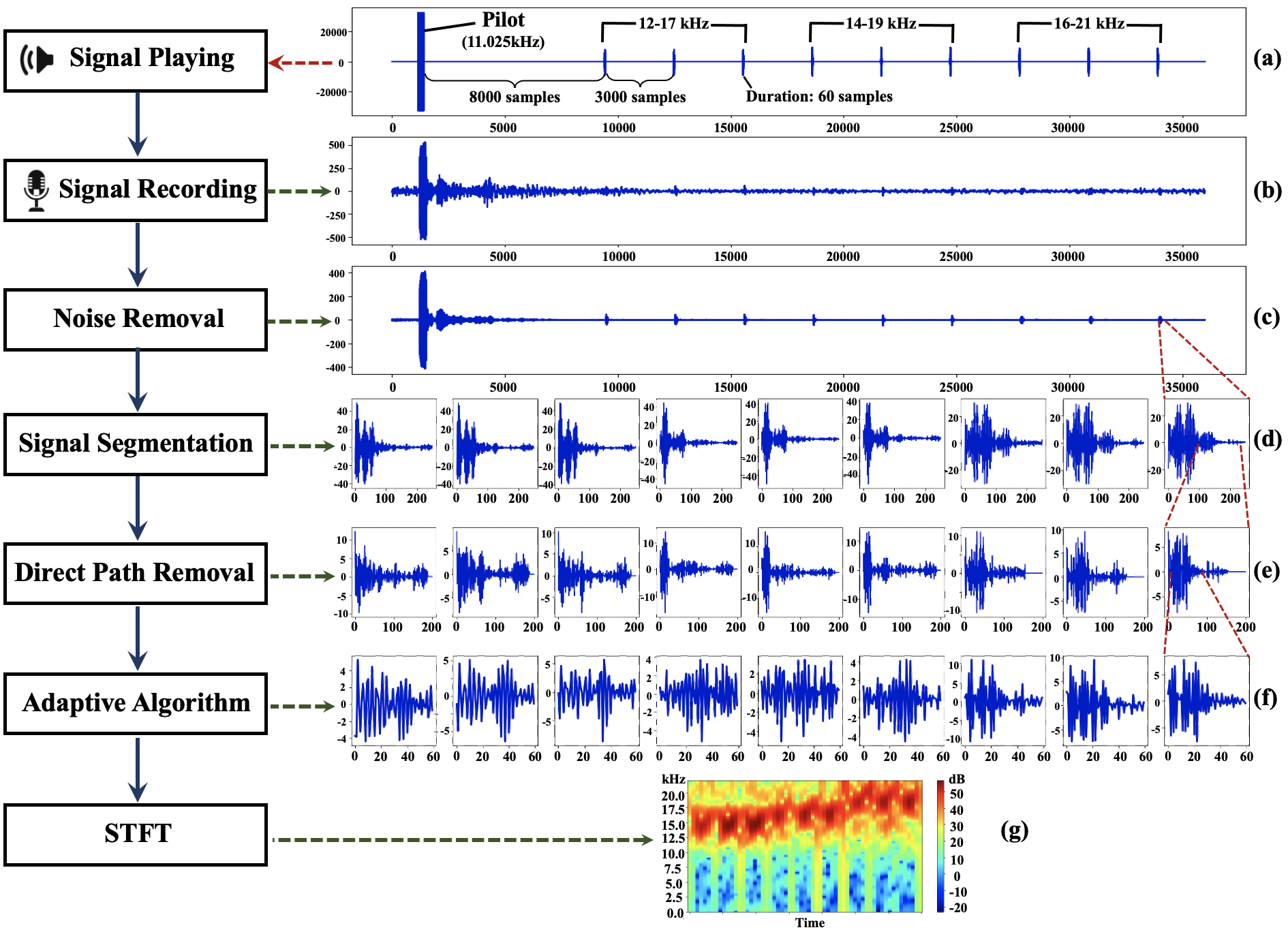}
\caption{Acoustic signal processing pipeline. The X and Y axes of sub-figure (a)-(f) indicate the sample point and amplitude. (a). The customized acoustic signal; (b). Recorded signal; (c). Signal after ambient noise removal; (d). Extracted nine chirps; (e). Chirps after direct path removal; (f). Extracted face region signals; (g). Spectrogram obtained by Short Time Fourier Transform (STFT).}
\label{signal_process}
\end{figure*}

\vspace{-0.2cm}
\section{Signal configuration and dataset acquisition}
\vspace{-0.0cm}
\subsection{Signal configuration} 
As Fig.~\ref{signal_process}(a) shows, the designed acoustic signal includes nine chirps with three different frequency sweep ranges (12-17 kHz, 14-19 kHz, and 16-21 kHz), thereby covering the frequency range from 12 to 21 kHz. The signal chirp at each group covers the corresponding frequency range in a linearly-increasing manner and repeats three times in the final emitted signal. This signal configuration comprises three benefits: high-quality anti-spoofing, high-level robustness, and minimum annoyance. Prior arts \cite{echoPrint2021, kong2022beyond} have demonstrated that high-frequency (HF) acoustic signals can capture rich and discriminative biometric patterns. Given that the surface geometric information of bonafide and 2D FPA should be distinctive in this work, we believe that the employed acoustic signal can achieve high-quality anti-spoofing. Besides, the acoustic signal is robust against ambient noise, which is usually \textless at 8 kHz \cite{zhou2021robust}. As the frequency of our signal ranges from 12 to 21 kHz, we can use a high-pass filter to eliminate the ambient noise's impacts.
Considering that good user experience in applications is vital, we propose minimizing the annoyance caused by the emitted acoustic signal. Although our signal frequency is high, the frequency range partially overlaps the audible frequency range of human beings (upper bound: 15-17 kHz on average~\cite{echoPrint2021}). Therefore, we increase the signal's peak-to-side ratio with a Hamming window~\cite{harris1978use} and control the play volume of the device to realize annoyance minimization and ensure good user experiences. As a result, the acoustic signal can achieve promising performance in high-accuracy FAS, robustness, and good user experience. 

\subsection{Dataset acquisition} 
Before data acquisition, we obtained the approval of the human ethics application from the Human and Artefacts Ethics Sub-Committee (Ref. no.: 1-2021-41-F). To facilitate developing an accurate and robust PAD system on mobile devices, we first build a large-scale two-modality database from 30 participants, henceforth referred to as \textbf{Echoface-Spoof}. During the data acquisition, the smartphone is held in front of the participant's face to ensure the quality of collected face pictures and acoustic signals. Other mobile devices can also be deployed without loss of generality. To accommodate the complex environments in practical scenarios and generalize to different mobile devices, we apply four commodity smartphones (\texttt{Samsung s9}, \texttt{Samsung s21}, \texttt{Samsung edge note}, and \texttt{Xiaomi Redmi7}) to collect acoustic and RGB data at different distances (25cm, 35cm, and 45cm), ambient noise levels (40dB, 60dB, and 70dB), and head poses for each participant.

Hardware degradation and intrinsic noise are inevitable in real-world applications. They can easily introduce quality distortions in captured pictures. In this vein, Echoface-Spoof incorporates a wide variety of common image distortion types, such as Gaussian blur, image color quantization with dither, JPEG compression, JPEG2000 compression, pink noise, and white noise. All variables above can affect the data quality of different modalities to some extent, which has been summarized in Table~\ref{variable}. 

We list existing 2D face spoofing databases in Table~\ref{FAS_databases}. Compared with previous datasets, Echoface-Spoof, for the first time, incorporates visual and auditory modalities for FAS. Furthermore, Echoface-Spoof uses different spoof mediums and devices with different resolutions during the data acquisition process to be more diverse. The proposed face spoofing database can shed light on the design of useful FAS systems and contribute to the progress in the community.

\begin{table}
  \caption{Impacts of different data collection variables.}
  \label{variable}
  \centering
  \renewcommand\arraystretch{1.15}
  \scalebox{1.0}{\begin{tabular}{c|c|c}
    \hline
    Variable & Vision & Acoustic \\
    \hline
    Device & \checkmark  & \checkmark \\
    \hline
    Distance & \checkmark  & \checkmark \\
    \hline
    Head pose & \checkmark  & \checkmark \\ 
    \hline
    Ambient noise & - & \checkmark  \\
    \hline
    Image distortion & \checkmark  & -\\
    \hline
\end{tabular}}
\end{table}

\begin{table*}
  \caption{Summary of existing 2D presentation attack databases}
  \label{FAS_databases}
  \centering
  \renewcommand\arraystretch{1.15}
  \scalebox{0.95}{\begin{tabular}{c|c|c|c|c|c}
    \hline
    Database & Release year & Modalities & \makecell[c]{\#Images or Videos\\(Live, spoof)} & Spoof medium & Acquisition device\\
    \hline
    \hline
    ZJU EyeBlink \cite{pan2007eyeblink} & 2007 & RGB & ( 80 , 100 ) & High-quality photo & Webcam (320$\times$240)\\
    \hline
    NUAA \cite{tan2010face} & 2010 & RGB & ( 5105 , 7509 ) & A4 paper & Webcam (640$\times$480)\\
    \hline
    \makecell[c]{IDIAP Print \\Attack \cite{anjos2011counter}} & 2011 & RGB & ( 200 , 200 ) & A4 paper & MacBook Webcam (320$\times$240)\\
    \hline
    CASIA FASD \cite{zhang2012face} & 2012 & RGB & ( 200 , 450 ) & \makecell[c]{iPad 1 (1024$\times$768)\\Printed photo} & \makecell[c]{Sony NEX-5 (1280$\times$720)\\USB camera (640$\times$480)\\Webcam (640$\times$480)}\\
    \hline
    \makecell[c]{IDIAP Replay \\Attack \cite{chingovska2012effectiveness}} & 2012 & RGB & ( 200 , 1000 ) & \makecell[c]{iPad 1 (1024$\times$768)\\iPhone 3GS (480$\times$320)} & \makecell[c]{MacBook Webcam (320$\times$240)\\Cannon PowerShot\\SX 150 IS (1280$\times$720)}\\
    \hline
    \makecell[c]{MSU-MFSD \cite{wen2015face}} & 2015 & RGB & ( 110 , 330 ) & \makecell[c]{iPad Air  (2048$\times$1536)\\iPhone 5s (1136$\times$640)\\A3 paper} & \makecell[c]{Nexus 5 (720$\times$480)\\MacBook (640$\times$480)\\Canon 550D (1920$\times$1088) \\iPhone 5s (1920$\times$1080)}\\
    \hline
    MSU-RAFS \cite{patel2015live} & 2015 & RGB & ( 55 , 110 ) & MacBook (1280$\times$800) & \makecell[c]{Nexus 5 (1920$\times$1080)\\ iPhone 6 (1920$\times$1080)}\\
    \hline
    \makecell[c]{IDIAP Multi-\\spectral-Spoof \cite{chingovska2016face}} & 2016 & RGB, Near-Infrared & ( 1689 , 3024 ) &A4 paper &u-Eye camera (1280$\times$1024)\\
    \hline
    MSU-USSA \cite{patel2016secure} & 2016 & RGB & ( 1140 , 9120) & \makecell[c]{MacBook (2880$\times$1800)\\Nexus 5 (1920$\times$1080)\\Tablet (1920$\times$1200)\\11$\times$8.5 in. paper} & \makecell[c]{Nexus 5 (3264$\times$2448)\\Cameras used to\\ capture celebrity photos}\\
    \hline
    OULU-NPU \cite{boulkenafet2017oulu} & 2017 & RGB & ( 1980 , 3960 ) & \makecell[c]{A3 paper\\Dell UltraSharp 1905FP\\ Display (1280$\times$1024)\\MacBook 2015 (2560$\times$1600)} & \makecell[c]{Samsung Galaxy S6 edge\\ HTC Desire EYE, OPPO N3\\MEIZU X5, ASUS Zenfore Selfie\\Sony XPERIA C5 Ultra Dual}\\
    \hline
    \makecell[c]{SiW (Spoofing\\in the Wild)\cite{liu2018learning}} & 2018 & RGB & ( 1320 , 3300 ) & \makecell[c]{Samsung Galaxy S8\\iPhone 7, iPad Pro\\ PC (Asus MB168B) screen} & \makecell[c]{Canon EOS T6\\ Logistech C920 webcam}\\
    \hline
    ROSE-YOUTU \cite{li2018unsupervised} & 2018 & RGB & 4225 & \makecell[c]{A4 paper\\Lenovo LCD (4096$\times$2160)\\Mac screen (2560$\times$1600)}&\makecell[c]{Hasee smartphone (640$\times$480)\\Huawei Smartphone (640$\times$480)\\iPad 4 (640$\times$480)\\iPhone 5s (1280$\times$720)\\ZTE smartphone (1280$\times$720)}\\
    \hline
    CASIA-SURF CeFA\cite{liu2021casia} & 2018 & \makecell[c]{RGB, Depth,\\ Infrared (IR)} & ( 7846 , 15692 ) & Paper, Silica gel mask&Intel Realsense\\
    \hline
    \makecell[c]{CUHK MMLab\\CelebA-Spoof \cite{CelebA-Spoof}} & 2020 & RGB & 625,537 & \makecell[c]{A4 paper\\Face mask, PC} & 10 sensors\\
    \hline
    Ambient-Flash \cite{di2020rethinking} & 2021 & \makecell[c]{RGB, additional \\light flashing} & ( 7503 , 7503 ) & \makecell[c]{Printed paper\\Digital screen} &\makecell[c]{Logitech C920 HD webcam\\ LenovoT430u laptop \\ webcam  (640$\times$480),  MotoG4}\\
    \hline
    \hline
    Echoface-Spoof (Ours) & 2024 & RGB, Acoustic & ( 82,715 , 166,637 ) & \makecell[c]{A4 paper\\iPad Pro (2388$\times$1668)\\iPad Air 3 (2224$\times$1668)}& \makecell[c]{Samsung  Edge Note (2560$\times$1440)\\Samsung Galaxy S9 (3264$\times$2448)\\Samsung Galaxy S21(4216$\times$2371)\\Xiaomi Redmi7 (3264$\times$2448)}\\
    \hline
\end{tabular}}
\end{table*}

\section{Methodology}
After data acquisition, we first design a signal  processing pipeline for acoustic fingerprint extraction. Then, we design a two-modality multi-scale neural network for accurate and flexible PAD.

\subsection{Overview of M$^3$FAS system}
Fig.~\ref{pipeline} depicts the pipeline of the M$^3$FAS system. The mobile device uses three built-in sensors to capture the raw data of presented bonafide/attack faces. The front camera captures the RGB visual picture. Meanwhile, the speaker emits a customized high-frequency signal, and the microphone collects the reflected acoustic signal. The captured raw data is subsequently sent to the data preprocessing module. For visual inputs, we use the face detector~\cite{king2009dlib} to crop the center face regions for accurate PAD. We further extract the acoustic fingerprint and convert it to the spectrogram map. Finally, the designed classification model conducts the decision-making.  

\subsection{Signal processing} 
\vspace{-1mm}
Fig.~\ref{signal_process} illustrates the acoustic signal processing pipeline of the M$^3$FAS system. The customized signal incorporates a pilot (11.025 kHz) and nine chirps with three frequency groups (12-17 kHz, 14-19 kHz, and 16-21 kHz). However, due to the inevitable ambient noise and hardware imperfections, the recorded signal is noisy, and the signal-noise ratio (SNR) is poor, as shown in Fig.~\ref{signal_process}(b). As such, we apply a high-pass filter to filter out the noise under 10 kHz, thereby eliminating the impact of ambient noise (\textless 8 kHz) meanwhile keeping the pilot (11.025 kHz) and chirps (12-21 kHz). 

After noise removal, we first locate the pilot of the signal in Fig.~\ref{signal_process}(c) by applying cross-correlation and locating the sample index of the highest peak. Then, given the pilot location, the interval between the pilot and the first chirp (8,000 samples), as well as the interval between two adjacent chirps (3,000 samples), we can readily locate the nine chirps in (c). We segment nine clips according to the obtained chirp locations, illustrated in Fig.~\ref{signal_process}(d). 

Each clip contains three components: direct path, target face echo, and background echo. Direct path denotes the signal directly transmitted from the speaker to the microphone, which should have the greatest energy and arrive at the microphone earliest among the three components. The target face echo and background echo are the signals ``reflected" by faces/spoof mediums and background objects ($e.g.$, walls), respectively. According to the travel distance, the face echo should have higher amplitudes and arrive at the microphone earlier than the background echo. In this vein, we can extract the target face echo from each clip since the three components are separable along the time dimension ($i.e.$, sample dimension). 

The direct path has the greatest energy and should be in the same shape as the emitted chirp. So we cross-correlate the original chirp with the corresponding clip and locate the most prominent peak ($i.e.$, the location of the direct path). We discard the direct path and send the processed clips in Fig.~\ref{signal_process}(e) to the adaptive algorithm. In this algorithm, we iteratively pass 60 samples in each clip to the original chirp to conduct cross-correlation, thereby can obtain nine face echo positions in one iteration. Ideally, face echo locations in the nine clips should be the same. Based on this theory, we pick the mean value of the nine positions with minimum standard deviation during the whole iteration process. Then, the mean value is taken as the face echo position of all nine clips. Once the face echo position is located, we extract 60 samples from each clip in (e) as the face echo. Finally, we concatenate the nine-face echoes along the time dimension and convert it to a spectrogram by adopting Short Time Fourier Transform (STFT).

\textcolor{black}{For the visual RGB signal, we first apply the popular \texttt{dlib} face detector~\cite{king2009dlib} to crop the face regions and subsequently normalize the resulting cropped face images to [-1, 1]. Finally, the preprocessed visual and audio data pair is then forwarded to the designed classification model.}

\begin{figure*}[ht]
\centering
\includegraphics[scale=0.60]{ 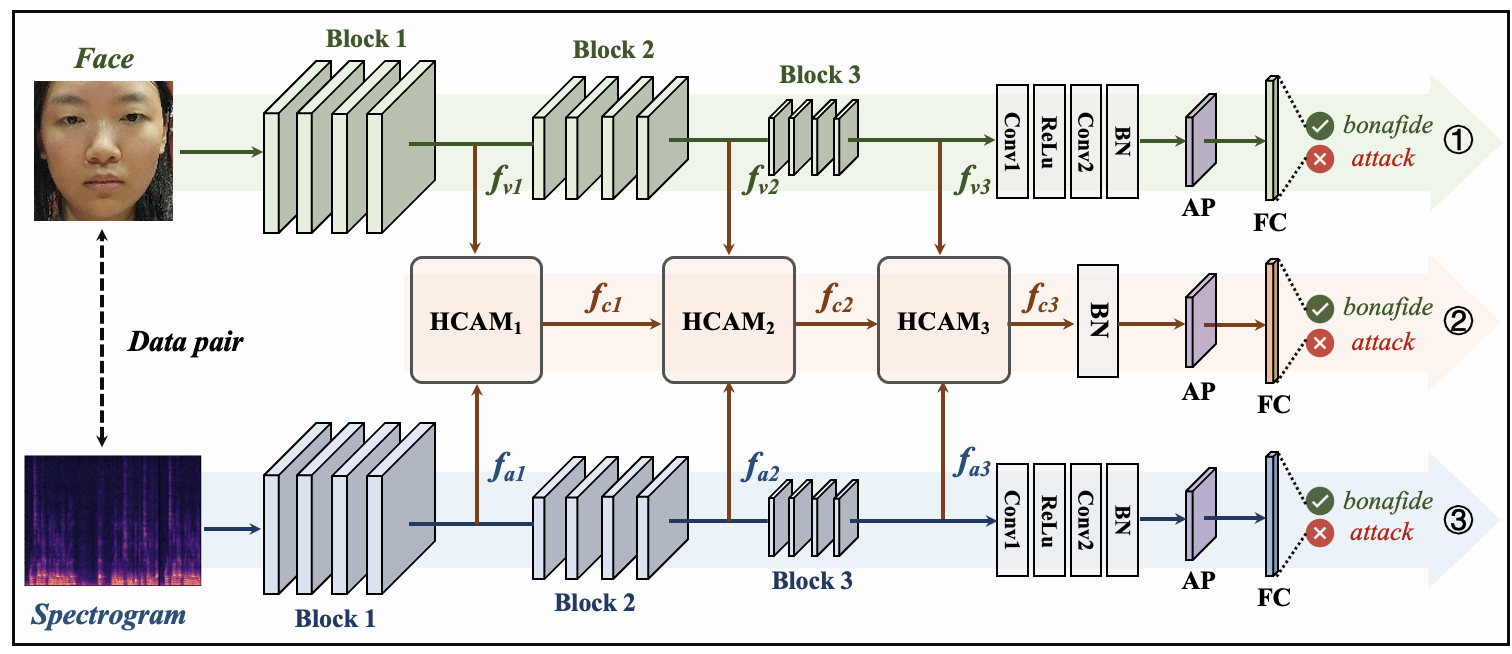}
\caption{Overview of the two-modality multi-scale FAS framework. BN: batch normalization; AP: average pooling; FC: fully-connected layer. In the training phase, the input data pair, including one cropped face picture and the corresponding spectrogram map, is synchronously fed forward to the vision and acoustic branch, each of which includes three blocks. The \textbf{H}ierarchical \textbf{C}ross-\textbf{A}ttention \textbf{M}odule (\textbf{HCAM}) is designed to interact better and align the vision feature \bm{$f_v$} and acoustic feature \bm{$f_a$}. The fused feature \bm{$f_c$} from previous \textbf{HCAM} will be hierarchically delivered to the next \textbf{HCAM} to accumulate features from different levels. Finally, three heads are designed to conduct decision-making respectively. In the testing stage, there are three different routes for inference: route \ding{172}, route \ding{173}, and route \ding{174}, thus enabling more flexible FAS in real-world applications. }
\label{framework}
\end{figure*}

\begin{figure*}[ht]
\centering
\includegraphics[scale=0.38]{ 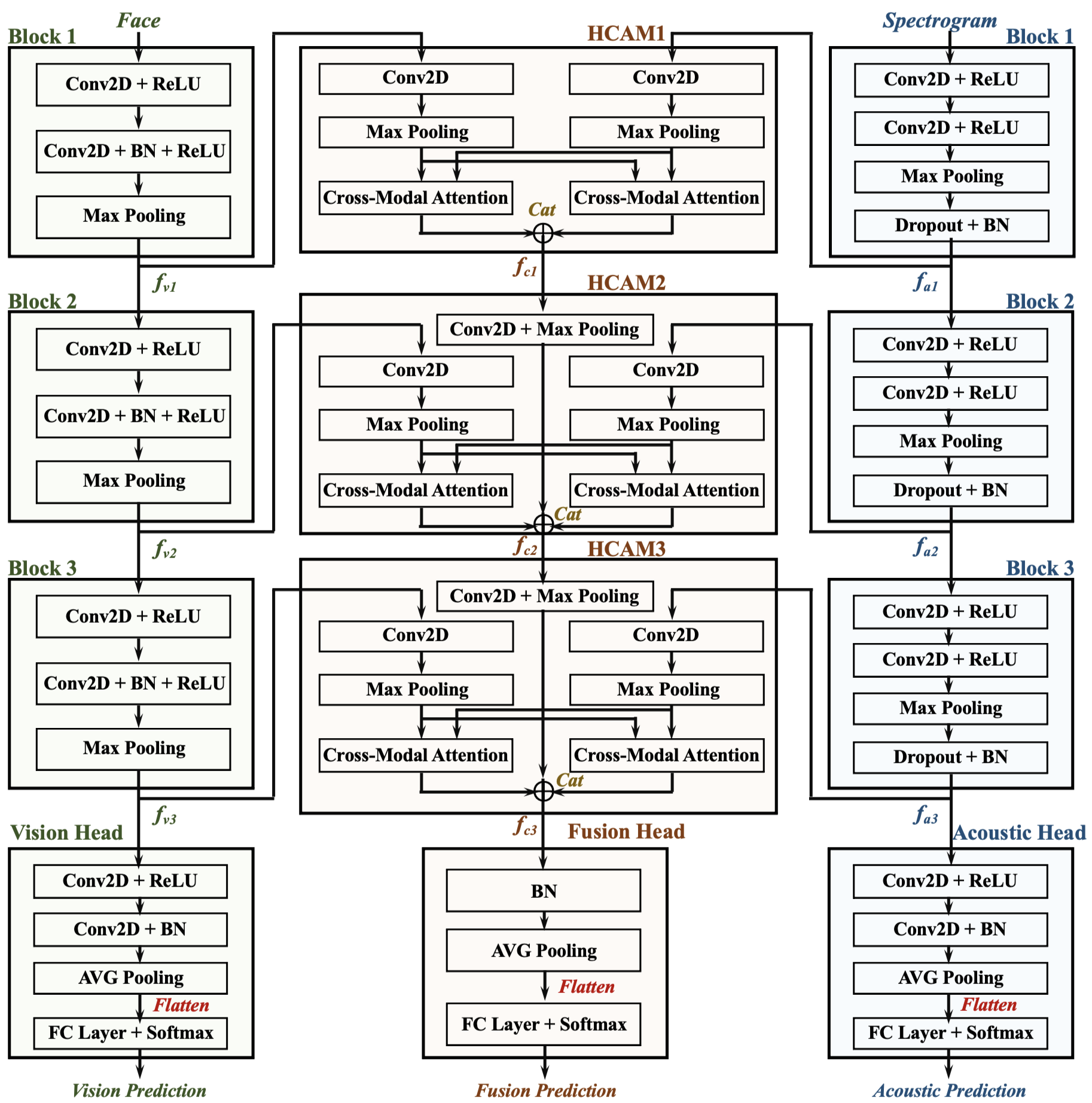}
\vspace{-4mm}
\caption{The details of M$^{3}$FAS model architecture. BN indicates Batch Normalization. FC Layer represents Fully-Connected Layer. Cat denotes concatenating different features along the channel dimension. }
\label{arch}
\end{figure*}

\begin{figure}[ht]
\centering
\includegraphics[scale=0.4]{ 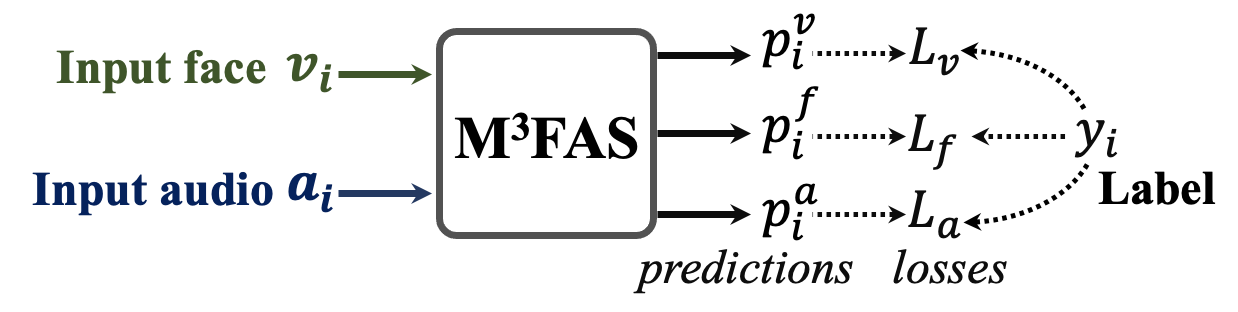}
\vspace{-5mm}
\caption{Details of loss calculations. $L_{f}$, $L_{a}$, and $L_{v}$ measure the disparities between the label $y_{i}$ and the predictions $p_{i}^{f}$, $p_{i}^{a}$, and $p_{i}^{v}$.}
\label{multihead}
\end{figure}


\subsection{Classification model}
\subsubsection{Two-modality multi-scale FAS framework}
We devise a two-modality multi-scale FAS framework for better interacting the visual and acoustic features and smartly aggregating the fused features from low, mid, and high levels. As Fig.~\ref{framework} depicts, the model takes face and spectrogram data pair as input and feeds them forward to the vision and acoustic branch. Each branch consists of three blocks, yielding three feature maps at different levels. \textcolor{black}{Each block in the designed framework consists of two convolution layers, two ReLu layers, one batch normalization layer, and one max. pooling layer.}
However, multi-modality optimization is a challenging problem since different modality tends to fit the data at different rates~\cite{kong2021appearance}. To mitigate this issue, we further design three hierarchical cross-attention modules (HCAM), equipping with cross-modality attention to better interact and align RGB and acoustic features.

In turn, HCAM takes advantage of multi-scale learning, aggregating hierarchical feature representations from low, mid, and high levels, thereby boosting PAD performance. Besides, three classification heads are placed at the top of the three branches. \textcolor{black}{As shown in Fig.~\ref{framework}, both vision and audio classification heads incorporate two convolution layers, one ReLu layer, one batch normalization layer, one average pooling layer, and one fully-connection layer, while the fusion head consists of one batch normalization layer and one average pooling layer.} 

\textcolor{black}{The M$^{3}$FAS framework takes a pair of face and spectrogram data as input. This input is then forwarded to the vision and acoustic branches, respectively. Fig.~\ref{arch} illustrates the details of the designed model architecture, with the left and right branches representing the vision and acoustic branches, respectively, while the middle branch serves as the fusion branch. Each vision or acoustic branch comprises three feature extraction blocks, each consisting of multiple 2D convolution layers, ReLu activation functions, and batch normalization (BN) layers. The incorporation of a Max Pooling layer in each block results in three feature maps at different levels. The extracted feature is simultaneously propagated to the next feature extraction block and the Hierarchical Cross-Attention Modules (HCAM). The fusion branch is composed of three HCAMs and one fusion classification head. HCAM is designed to hierarchically aggregate the multimodal features. For example, {$\rm {HCAM_{2}}$} takes the visual feature $f_{v1}$, auditory feature $f_{a1}$, and the fused feature from the previous HCAM $f_{c1}$ as inputs. These features first undergo one convolution layer and one max pooling layer. The processed visual and acoustic features are then fed into two cross-modal attention modules, generating two attended features, which are concatenated with the processed $f_{c2}$ along the channel dimension. The resulting $f_{c2}$ effectively aggregates multimodal features and the fused features from the previous layer, which are then forwarded to the next HCAM. Three classification heads are positioned on top of each branch, producing three predictions simultaneously.  Three binary cross-entropy losses between the label and the predictions are combined to supervise the training process.}

In the training phase, we adopt the joint training strategy. \textcolor{black}{As each data pair consists of a face image and its corresponding acoustic signal, both sharing an identical label. In our proposed multi-head learning strategy, we leverage the common label to individually supervise the training of the vision, acoustic, and fusion branches .} Three binary cross-entropy (BCE) losses are combined to train the whole model. The adopted joint training strategy comprises two merits: (1). It enables more flexible PAD in the inference stage; (2). It effectively mitigates the overfitting problem and boosts the PAD performance for a single branch, compared with the separate training.
The objective function is given as follows:
\begin{equation}
     L_{total}= L_{f} + \alpha(L_{v} + L_{a}),
\end{equation}
where $L_{f}$, $L_{v}$, and $L_{a}$ are BCE loss functions for fusion, vision, and acoustic branches. $\alpha$ weighs different loss components. \textcolor{black}{As shown in Fig.~\ref{multihead}, $L_{f}$, $L_{a}$, and $L_{v}$ measure the disparities between the label $y_{i}$ and the predictions $p_{i}^{f}$, $p_{i}^{a}$, and $p_{i}^{v}$ for the fusion, acoustic, and vision branches, respectively.}

\begin{figure}[ht]
\centering
\includegraphics[scale=0.4]{ 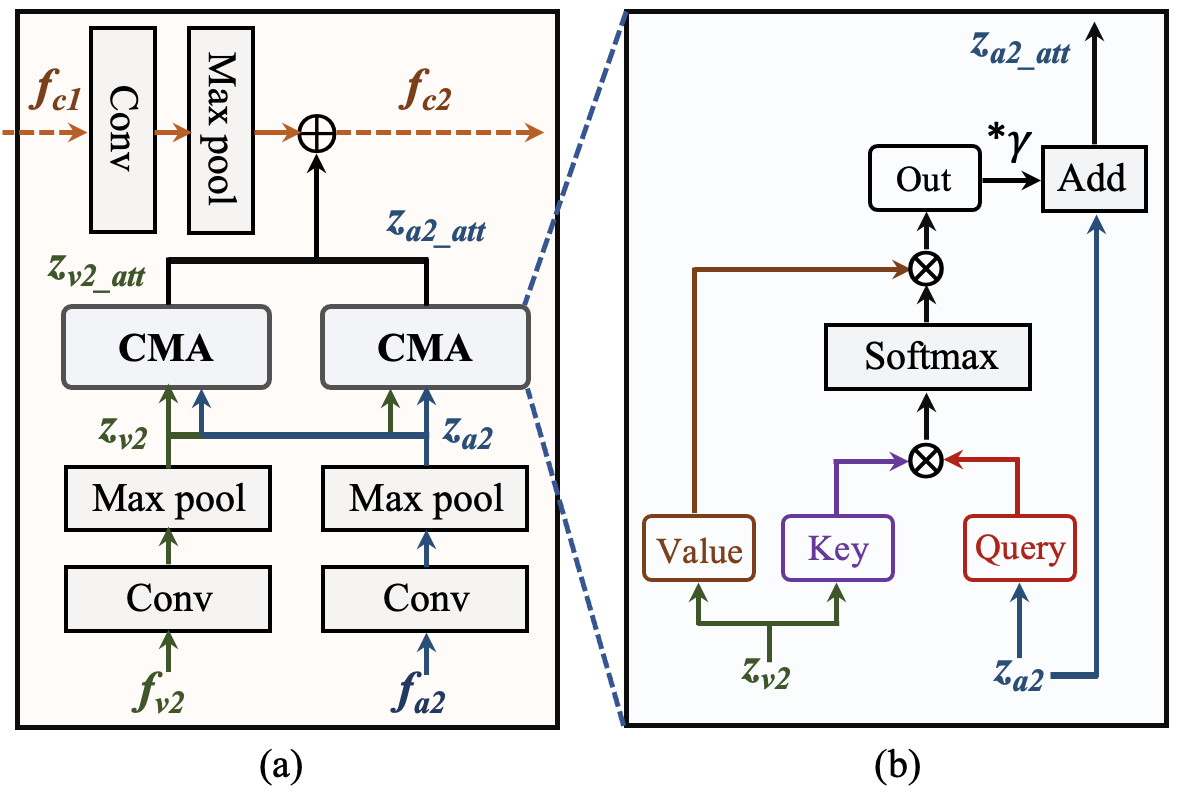}
\vspace{-5mm}
\caption{Illustration of the \textbf{H}ierarchical \textbf{C}ross-\textbf{A}ttention \textbf{M}odule (\textbf{HCAM}). Herein, we depict {$\rm \bf{HCAM_{2}}$} to illustrate the network architecture, where $\bigoplus$ denotes the concatenation operator. \bm{$f_{c1}$} represents the feature from the previous HCAM. The right figure dedicates the details of the cross-modality attention \textbf{CMA} module, where $\gamma$ is a trainable parameter, and $\bigotimes$ denotes the matrix multiplication. 
}
\label{HCAM}
\end{figure}

\textcolor{black}{During testing, if both modality data are available, we directly take the prediction result of the fusion head as the final decision due to its superior detection accuracy to other two branches. However, in cases where the visual modality is unavailable, the fusion and visual branches remain inactive, and the auditory output will be taken as final decision. Similarly, if the auditory modality is absent, the final decision is determined by the visual output.}


\subsubsection{Hierarchical cross-attention module (HCAM)}
In this section, we dedicate the details of HCAM. \textcolor{black}{In comparison to the attention mechanism discussed in~\cite{kong2022beyond}, HCAM introduces cross-modality attention modules that better align and interact spoofing features from visual and auditory modalities. Furthermore, HCAM proposes to hierarchically accumulate the multimodal features across low, middle, and high levels, thereby enhancing the final PAD accuracy and robustness.} Fig.~\ref{HCAM} (a) takes {$\rm \bf{HCAM_{2}}$} as the showcase. $f_{v2}$ and $f_{a2}$ are features extracted from Block 2 of the vision and acoustic branch. The features first go through one convolutional layer and one max-pooling layer, yielding $z_{v2}$ and $z_{a2}$. Inspired by the recent success of the attention mechanism in natural language processing and computer vision, we exploit two cross-modality attention (CMA) modules to align the features of the two modalities better. We illustrate the architecture details of CMA in  Fig.~\ref{HCAM} (b). Cross-modality attention (CMA) can be formulated as follows: 

\begin{figure}[ht]
\centering
\includegraphics[scale=0.28]{ 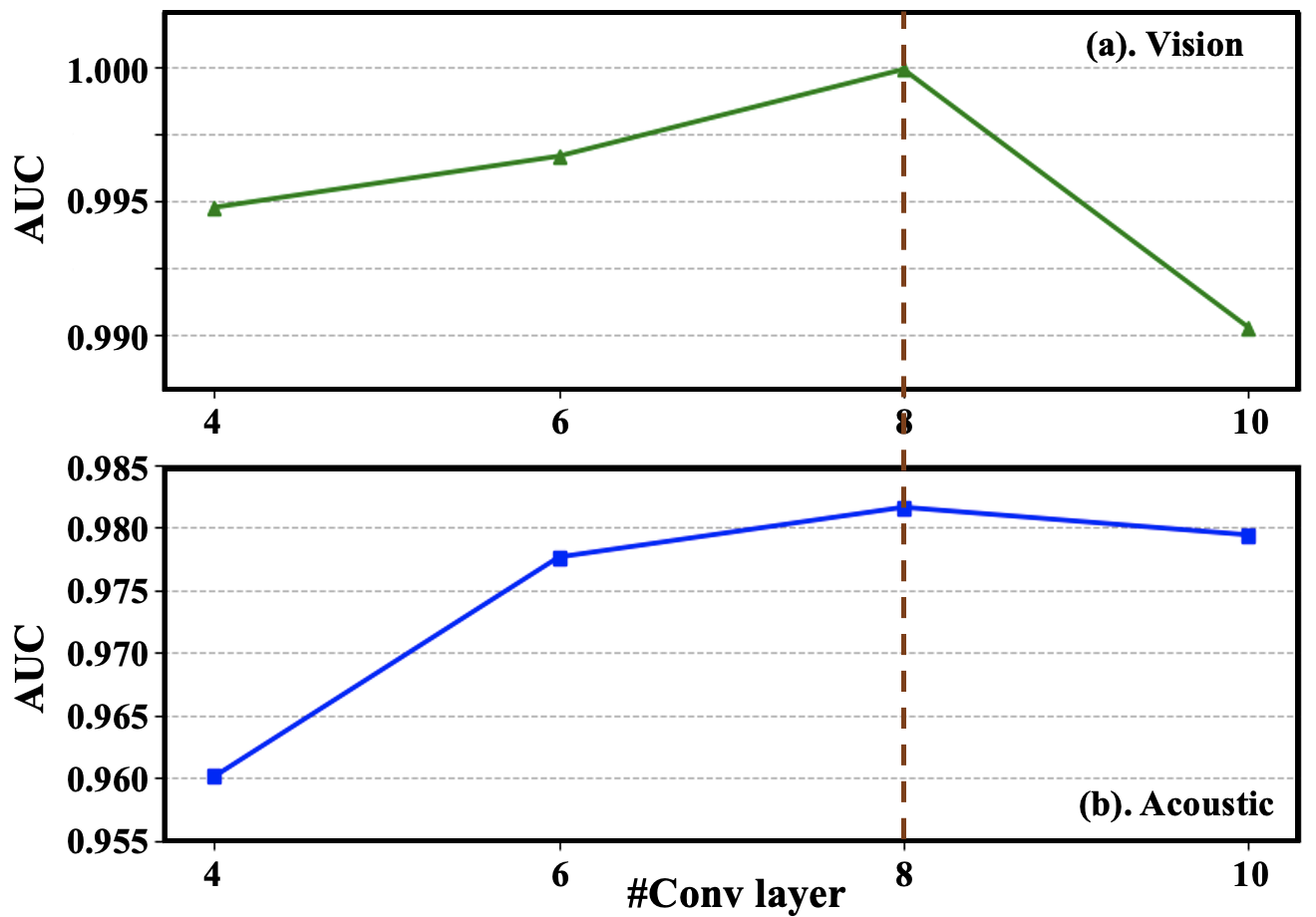}
\vspace{-3mm}
\caption{AUC score versus convolutional numbers for (a). vision branch, and (b). the acoustic branch under the cross-subject setting.}
\label{conv_layer}
\end{figure}

\begin{figure}[ht]
\centering
\includegraphics[scale=0.32]{ 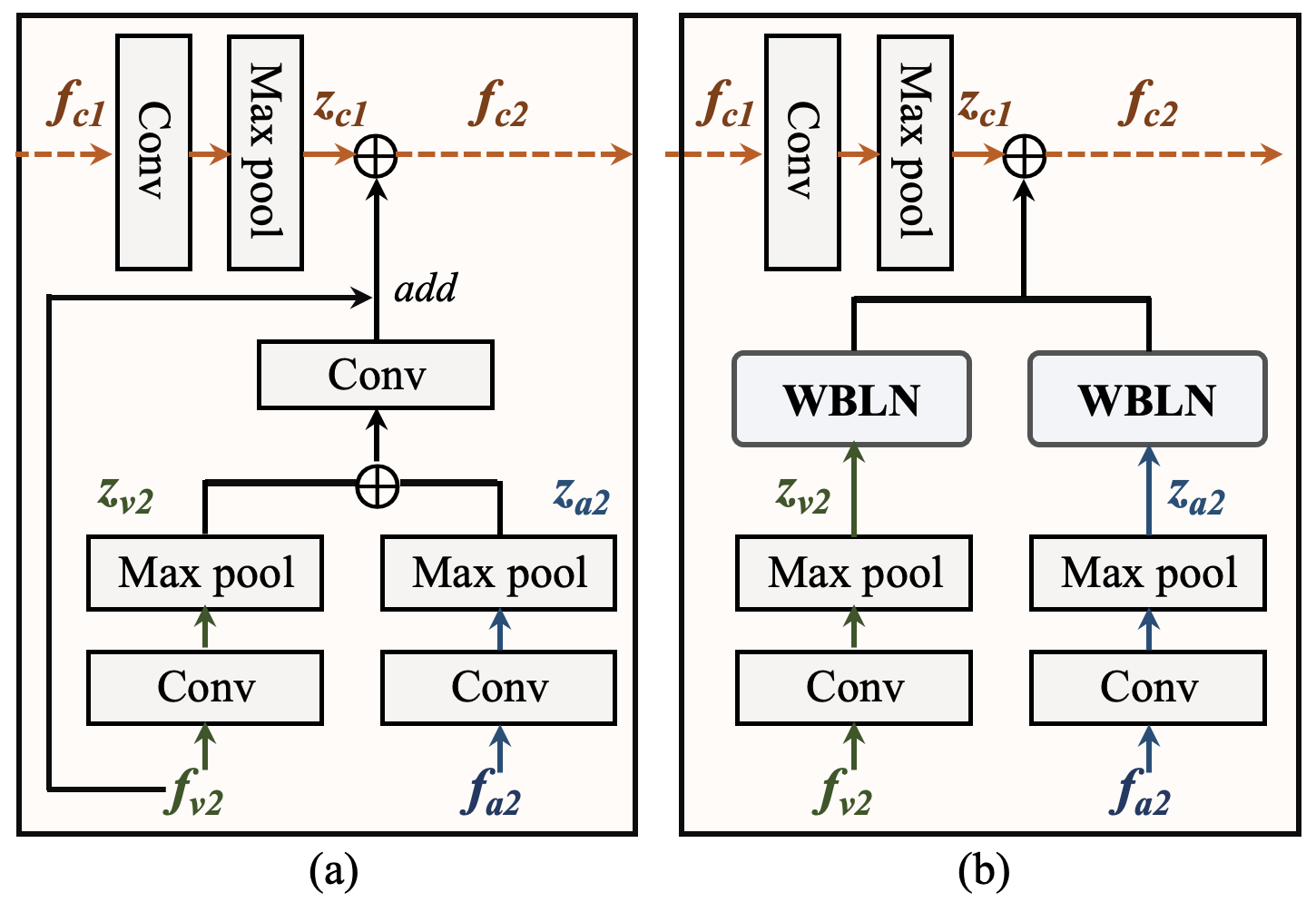}
\caption{Two-modality feature fusion strategies: (a). Residual feature fusion: $f_{c2}$=$z_{c1}$$\oplus$[$f_{v2}$+Conv($z_{v2}$$\oplus$$z_{a2}$))]; (b). Weighted batch  and layer normalization: $f_{c2}$=$z_{c1}$$\oplus$WBLN($z_{v2}$)$\oplus$WBLN($z_{a2}$), where WBLN(f)=(1-$\theta$)*Batchnorm(f)+$\theta$*Layernorm(f).}
\label{fusion_strategy}
\end{figure}

\begin{equation}
     z_{a\_att}= z_{a} + \gamma{\rm softmax}(\frac{z_{a}W_{Q}(z_{v}W_{K})^\top} {\sqrt{d}})(z_{v}W_{V}),
\end{equation}
\begin{equation}
     z_{v\_att}= z_{v} + \gamma{\rm softmax}(\frac{z_{v}W_{Q}(z_{a}W_{K})^\top} {\sqrt{d}})(z_{a}W_{V}),
\end{equation}
where $W_{Q}$, $W_{K}$, and $W_{V}$ are learnable parameters of linear projection layers. We set a learnable coefficient $\gamma$ to weigh the learned attention component smartly.

Then, we concatenate the attended features as well as the feature $f_{c1}$ from {$\rm \bf{HCAM_{1}}$} along the channel dimension, and the produced feature $f_{c2}$ is subsequently forwarded to {$\rm \bf{HCAM_{3}}$}. In summary, the design philosophy of HCAM promises two merits: (1). It takes advantage of multi-scale learning, aggregating hierarchical feature representations from low, mid, and high levels, thereby boosting the final PAD performance; (2). It better aligns RGB and acoustic features, thus effectively mitigating the overfitting problem of multi-modality optimization. 

\vspace{-4mm}
\section{Experiments}
In this section, we conduct ablation studies to demonstrate the effectiveness of the designed network architecture. We evaluate our model in terms of security, robustness, and flexibility under a wide variety of experimental settings.

\vspace{-3mm}
\subsection{Implementation Details} The model is implemented by Pytorch~\cite{paszke2019pytorch}. We use Adam optimizer \cite{kingma2014adam} with {$\beta_{1}$}=0.9 and {$\beta_{2}$}=0.999 to train the designed model on 1 RTX 2080Ti GPU with batch size 256. The learning rate and weight decay are set as 1e-4 and 1e-5,  respectively. The framework is trained for 100 epochs and validated at the end of every epoch, and the checkpoint with the best validating HTER value is picked for testing.

\subsection{Evaluation Metrics} 
Following the literature, we use AUC and HTER as the evaluation metrics. Besides, we also report ACC and EER in this paper for better interpreting the experimental results. 

\noindent\textbf{Area Under Curve (AUC)} measures the area under the Receiver Operating Characteristic (ROC) curve. 

\noindent\textbf{Half Total Error Rate (HTER)} calculates the average of the False Acceptance Rate (FAR) and False Reject Rate (FRR). A lower HTER value indicates better PAD performance:
\begin{equation}
    \label{hter}
 HTER = \frac{FAR + FRR} {2} = \frac{1} {2}\Big(\frac{FP} {TN + FP} + \frac{FN} {TP + FN}\Big)
\end{equation}

\noindent\textbf{Accuracy (ACC)} directly reflects the probability that one query can be correctly classified: 
\begin{equation}
    \label{acc}
 ACC = \frac{TP + TN} {TP + FP + TN + FN} 
\end{equation}
where TP, TN, FP, and FN indicate true positive, true negative, false positive, and false negative, respectively.  

\noindent\textbf{Equal Error Rate (EER)} measures the optimal point for the trade-off between false positive and false negative. It is the False Positive Rate (FPR) that equals the True Positive Rate (TPR). 

\subsection{Ablation study}
Before proceeding to comparative experiments, we first conduct the ablation study to validate the effectiveness of the designed network architectures, involved components, and the proposed joint training strategy. Besides, we also investigate the impacts of different loss weight values to weigh the loss components better and improve the final detection performance. We conduct most ablation experiments under the cross-device setting.


\begin{table}
  \caption{Face liveness detection performance with different feature fusion strategies.}
  \label{abl1}
  \centering
  \renewcommand\arraystretch{1.15}
  \scalebox{1.0}{\begin{tabular}{c|c|c|c|c}
    \hline
     Strategy & AUC(\%) $\uparrow$ & ACC(\%) $\uparrow$ & HTER(\%) $\downarrow$ & EER(\%) $\downarrow$\\
    \hline
    Cat & 99.2 & 90.3 & 13.7 & 3.2\\
    \hline
    AVG & \textbf{99.7} & 88.4 & 16.9 & 1.8 \\
    \hline
    RES & 99.5 & 91.9 & 11.8 & 2.2\\ 
    \hline
    WBLN & 99.2 & 87.9 & 15.5 & 2.6\\
    \hline
    CA & \textbf{99.7} & \textbf{92.8} & \textbf{10.3} & \textbf{1.4} \\
    \hline
\end{tabular}}
\end{table}

\subsubsection{Neural architecture search (NAS)}
Considering that a new database is proposed in this work, we first conduct a brief neural architecture search to determine the backbones of the two branches. Fig.~\ref{conv_layer} depicts the AUC values versus convolution layer numbers for vision and acoustic branches under the cross-subject evaluation setting. Finally, the layer number is determined as 8 for both branches.

\subsubsection{Effectiveness of feature fusion strategy}
Combining features of different modalities properly is challenging because their distributions differ dramatically in feature space. We have tried five different fusion methods to investigate the best strategy for fusing vision and acoustic features and report the detection performance under the cross-device evaluation in Table~\ref{abl1}. Cat indicates that the two features are concatenated along the channel dimension, and AVG means conducting an average operation on the input vision and acoustic features. RES and WBLN~\cite{yu2022benchmarking} have been illustrated in Fig.~\ref{fusion_strategy}, where RES aims at learning the fusion feature in a residual fashion, and WBLN conducts weighted batch and layer normalization over the input features. Inspired by the recent success of the attention mechanism, we also use cross-attention (CA) in this feature fusion experiment. From Table~\ref{abl1}, we observe that CA achieves the best detection performance. As such, we use cross-attention to align and fuse the input vision and acoustic features in our framework.

\begin{table}
  \caption{PAD performance of the two-branch framework equipped with HCAM modules under the cross-device setting.}
  \label{abl_ms}
  \centering
  \renewcommand\arraystretch{1.15}
  \scalebox{0.78}{\begin{tabular}{c|c|c|c|c|c|c}
    \hline
     {$\rm \bf{HCAM_{1}}$} & {$\rm \bf{HCAM_{2}}$} & {$\rm \bf{HCAM_{3}}$} & AUC(\%) $\uparrow$ & ACC(\%) $\uparrow$ & HTER(\%) $\downarrow$ & EER(\%) $\downarrow$\\
    \hline
    - & - & \checkmark & 99.2 & 90.0 & 14.6 & 2.2 \\
    \hline
    - & \checkmark & \checkmark & 98.5 & 90.3 & 14.1 & 3.5 \\ 
    \hline
    \checkmark & \checkmark & \checkmark & \textbf{99.7} & \textbf{92.8} & \textbf{10.3} & \textbf{1.4} \\
    \hline
\end{tabular}}
\end{table}

\begin{figure}[ht]
\centering
\includegraphics[scale=0.27]{ 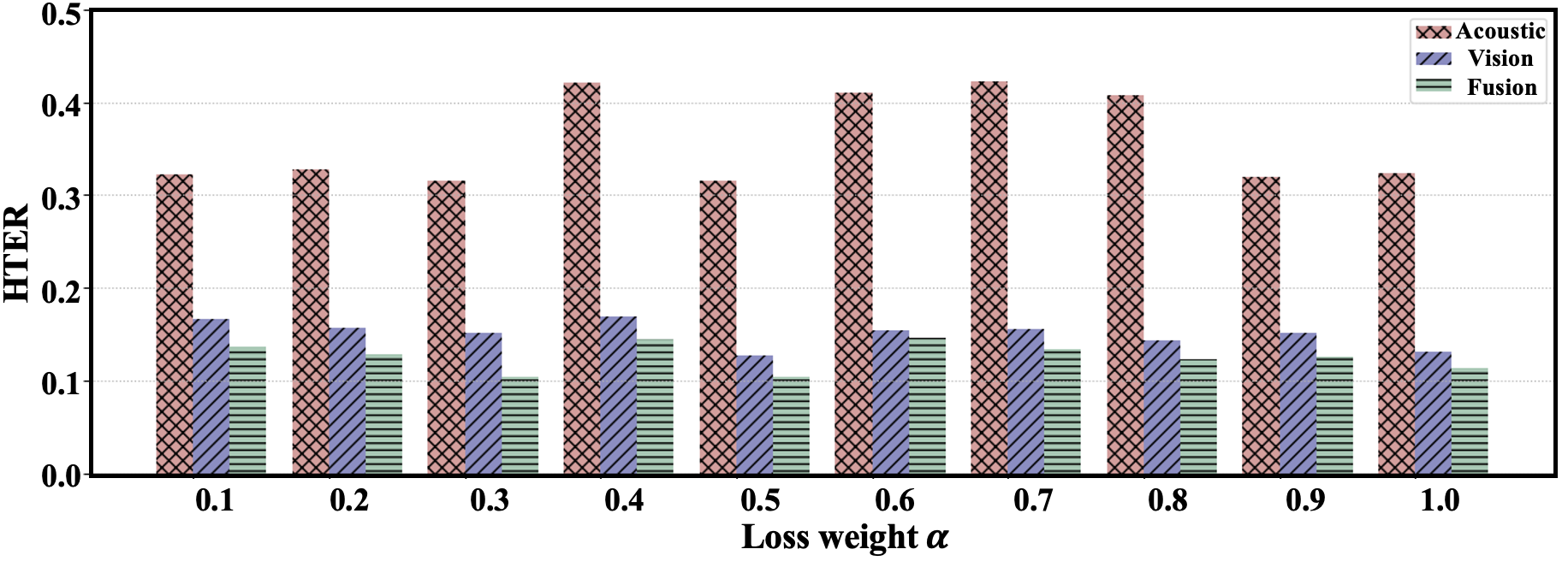}
\vspace{-3mm}
\caption{Impacts of different loss weight values on the fusion, vision, and acoustic heads.}
\label{Lossweight}
\end{figure}

\begin{figure}[ht]
\centering
\includegraphics[scale=0.26]{ 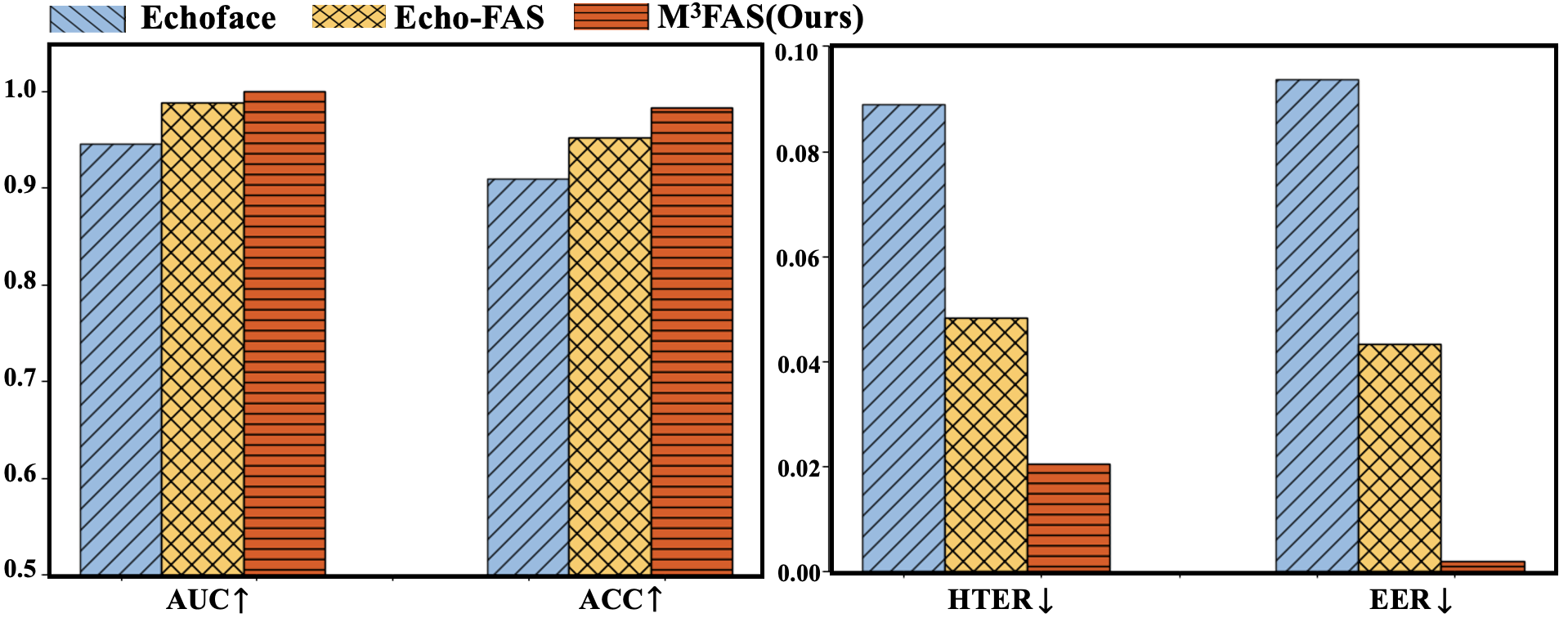}
\vspace{-0.3cm}
\caption{Comparison with prior arts: Echoface\cite{chen2019echoface} and Echo-FAS \cite{kong2022beyond}.}
\label{sota_methods}
\end{figure}

\subsubsection{Effectiveness of multi-scale learning scheme} 
This paper uses three HCAMs to conduct the hierarchical feature aggregation from three scales. The multi-scale learning scheme takes advantage of low-, mid-, and high-level feature representations, thus contributing to the final performance boosts. To verify this assumption, we report the detection results with different HCAMs in Table~\ref{abl_ms}. It can be readily observed that more aggregated features lead to better PAD performance, demonstrating the effectiveness of the proposed multi-scale learning scheme.

\subsubsection{Effectiveness of joint training strategy}
As Fig.~\ref{framework} shows, we propose a novel joint training strategy in the M$^3$FAS system. This strategy enables a more flexible PAD in the testing phase. On the other hand, we find it can effectively mitigate the overfitting problem and improve the detection accuracy of the single branch. We report the face liveness detection results of separate training and joint training strategies under both cross-device and cross-subject evaluations in Table~\ref{joint_sep}. For each modality, the overall detection performance of joint training is superior to the separate training under both settings, demonstrating the efficiency of the proposed joint training strategy.


\subsubsection{Impacts of different loss weights}
To determine the value of loss wight $\alpha$ in Eqn.~(1), we measure the HTER values of three predictions versus different $\alpha$ under the cross-device evaluation setting. Fig.~\ref{Lossweight} depicts the impacts of different $\alpha$ values on the predictions produced by the fusion, vision, and acoustic heads. We can see the fusion results are consistently superior to the other two heads regardless of the vary of $\alpha$, which demonstrates the effectiveness of the designed network architecture from another point of view. Therefore, the $\alpha$ value is determined as 0.5 to appropriately weigh different loss components in Eqn.~(1).

\subsubsection{Discussion}
We have conducted extensive and rigorous ablation experiments to search the network architecture and evaluate the impacts of different loss weight values. On the other hand, we further verify the benefits of the involved HCAM components, the multi-scale learning scheme, and the joint training strategy. In the rest of the section, we aim to verify the efficiency of the devised M$^3$FAS system from security, robustness, and flexibility perspectives.

\subsection{Security evaluations of M$^3$FAS} 
The devised M$^3$FAS system must be secure enough to safeguard face authentication systems against various face presentation attacks accurately. In real-world applications, FAS systems should be directly deployed to unforeseen identities. As such, we train the classification model on 25 subjects and test it on the data of the other five subjects. Moreover, training and testing data are captured by the same mobile device. Thanks to the joint training strategy, we can report the face liveness detection performance of vision, acoustic, and fusion heads in Table~\ref{intra_device}. The average results are highlighted in gray, and the bold numbers specify the best average detection results. Two conclusions can be easily drawn: (1). vision and acoustic branch can achieve promising performance, and the fusion model, taking advantage of visual texture and acoustic geometric spoofing cues, achieves the best average detection results; (2). M$^3$FAS achieves 99.9\% AUC and 98.3\% ACC classification scores, which are satisfactory for deployment. The auditory modality can achieve outstanding detection solely, and it also plays a critical role in boosting the final detection performance of the cross-modal fusion framework.

Furthermore, we compare the proposed M$^3$FAS system with prior mobile FAS systems in Fig.~\ref{sota_methods}. It should be noted that it is not an ideally fair comparison as the proposed method has been fed with more data (both RGB and acoustic). Nevertheless, again, M$^3$FAS achieves the best detection results. 


\begin{table}
  \caption{Face liveness detection results of separate training and joint training strategies under cross-device and cross-subject settings.}
  \label{joint_sep}
  \centering
  \renewcommand\arraystretch{1.15}
  \scalebox{0.85}{\begin{tabular}{c|c|c|c|c|c}
    \hline
    Cross-device & Modality & AUC(\%)$\uparrow$ & ACC(\%)$\uparrow$ & HTER(\%)$\downarrow$ & EER(\%)$\downarrow$\\
    \hline
    \hline
    \multirow{2}*{\textbf{Separate}}& Vision  & 94.2 & 90.2 & 13.7 & 12.0 \\
	\cline{2-6}
	~ & Acoustic & 85.1 & 61.6 & 41.1 & 20.3\\
	\cline{2-6}
    \hline
    \hline
    \multirow{2}*{\textbf{Joint}} 
	~ & Vision  & 96.9 & 91.2 & 12.7 & 5.4\\
	\cline{2-6}
        ~ & Acoustic & 85.0 & 69.0 & 31.5 & 21.1 \\
        \cline{2-6}
        \hline
    \hline
    Cross-subject & Modality & AUC(\%)$\uparrow$ & ACC(\%)$\uparrow$ & HTER(\%)$\downarrow$ & EER(\%)$\downarrow$\\
    \hline
    \hline
    \multirow{2}*{\textbf{Separate}}& Vision & 99.9 & 97.7 & 3.1 & 0.8\\
	\cline{2-6}
	~ & Acoustic & 97.8 & 93.7 & 6.4 & 6.0 \\
	\cline{2-6}
    \hline
    \hline
    \multirow{2}*{\textbf{Joint}} 
	~ & Vision  & 99.9 & 97.8 & 2.8 & 0.4 \\
	\cline{2-6}
        ~ & Acoustic & 98.1 & 93.8 & 7.1 & 5.4 \\
        \cline{2-6}
        \hline
\end{tabular}}
\end{table}


\subsection{Robustness evaluations of M$^3$FAS} 
Learning-based FAS models are prone to overfitting issues, resulting in limited generalization capability. Moreover, current FAS systems are more vulnerable to presentation attacks from unseen environments. Therefore, evaluating the robustness of the proposed M$^3$FAS system is of utmost importance. We conduct extensive cross-domain experiments in this section, including cross-device evaluation, cross-environmental variable evaluation, and cross-distortion evaluation.  


\subsubsection{Generalization to unseen devices}
\begin{figure*}[ht]
\centering
\includegraphics[scale=0.41]{ 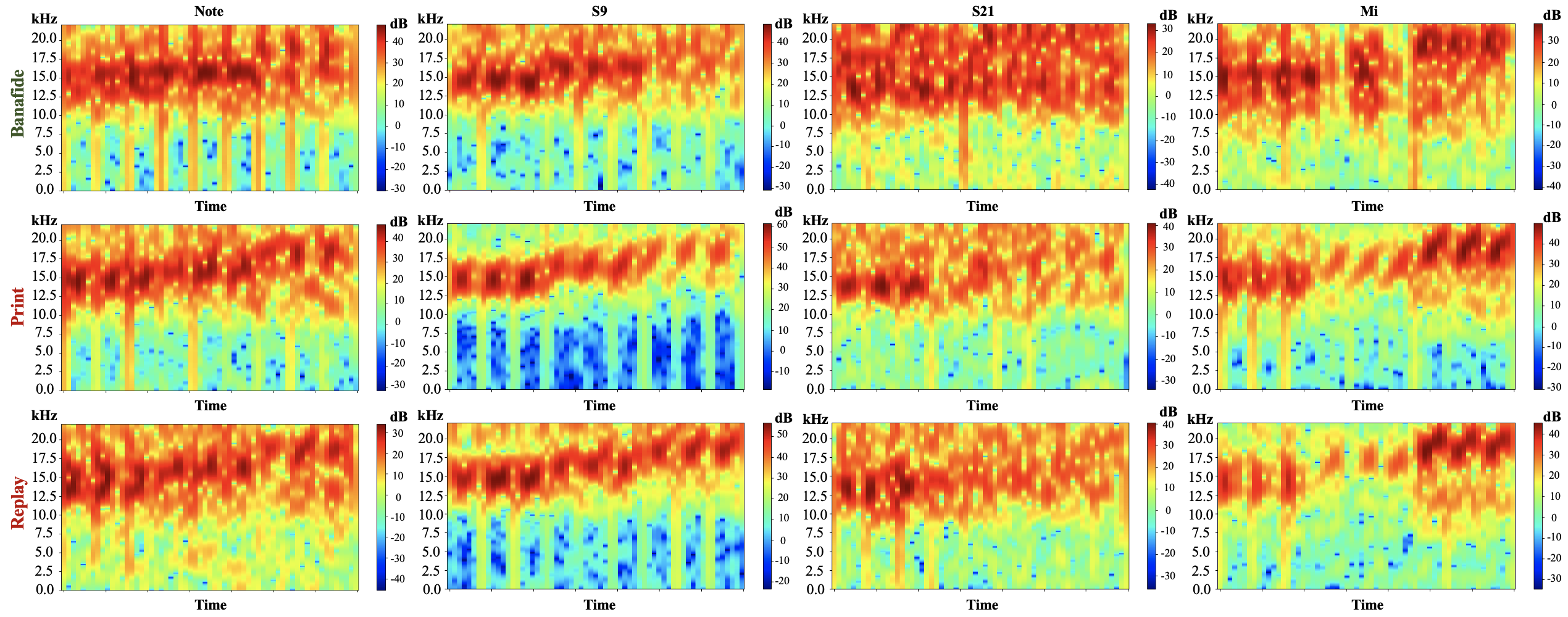}
\vspace{-8mm}
\caption{The visualization of spectrograms for different data acquisition devices. The top, middle, and bottom rows correspond to bonafide, print attack, and replay attack, respectively.}
\label{spect}
\end{figure*}

Cross-device FAS is a common setting in real-world applications, as the trained model can be directly deployed on unseen devices in a plug-and-play fashion. In this experiment, we train the model on three devices, test it on the other device, and report the results in Table~\ref{cross_device}, where ``Device" specifies the testing device. We still keep the training and testing subjects have no overlap in this evaluation. Compared with the Intra-device results in Table~\ref{intra_device}, face liveness detection performance of vision, acoustic, and fusion heads all suffer from a certain degradation due to the data distribution gaps among different devices. Specifically, we found that the acoustic branch suffers a dramatic performance drop under the cross-device setting because the hardware diversity (speaker and microphone) introduced by manufacture imperfection will cause different frequency responses and non-uniform intrinsic noises~\cite{gao2022device, liu2022soundid} among different mobile devices. \textcolor{black}{Chen et al. \cite{chen2021camera} demonstrated that face images acquired by different cameras have distinct spoofing patterns. And \cite{kong2022beyond} demonstrated that the frequency responses of different smartphones differ dramatically. Consequently, these hardware discrepancies can indeed cause severe domain gaps among data acquisition devices, resulting in FAS performance drops in the cross-device evaluation \cite{jain2021biometrics}.}
We visualize the spectrogram maps of the four smartphones in Fig.~\ref{spect}. It can be observed that the spectrogram maps of different devices are discriminative. The acoustic signal can still play a complementary role to the visual modality and boost the detection performance of the fusion head. The fusion head can achieve a promising 99.7\% AUC and 92.8\% ACC cross-device detection performance, demonstrating the robustness of the multi-modal FAS system. Compared to the visual modality, the AUC score has been improved by 2.8\% (from 96.9\%$\rightarrow$99.7\%), demonstrating the effectiveness of the auditory modality.

\begin{table}
  \caption{Cross-subject face liveness detection results.}
  \label{intra_device}
  \centering
  \renewcommand\arraystretch{1.15}
  \scalebox{0.90}{\begin{tabular}{c|c|c|c|c|c}
    \hline
    Modality & Device & AUC(\%)$\uparrow$ & ACC(\%)$\uparrow$ & HTER(\%)$\downarrow$ & EER(\%)$\downarrow$\\
    \hline
    \hline
    \multirow{5}*{\textbf{Vision}} & Note & 99.9 & 93.8 & 9.2 & 1.1 \\
	\cline{2-6}
	~ & S9 & 99.9 & 98.5 & 1.1 & 0.1 \\
	\cline{2-6}
        ~ & S21 & 99.9 & 99.8 & 0.2 & 0.2 \\
        \cline{2-6}
        ~ & Mi  & 99.9 & 99.0 & 0.7 & 0.2 \\
        \cline{2-6}
        ~ & \cellcolor{lightgray} AVG & \cellcolor{lightgray}\textbf{99.9} & \cellcolor{lightgray}97.8 & \cellcolor{lightgray}2.8 & \cellcolor{lightgray}0.4\\
    \hline
    \hline
    \multirow{5}*{\textbf{Acoustic}} & Note & 97.1 & 91.4 & 10.1 & 7.1 \\ 
	\cline{2-6}
	~ & S9  & 99.5 & 95.8 & 5.8 & 3.0 \\
	\cline{2-6}
        ~ & S21  & 96.4 & 90.9 & 9.4 & 9.0 \\
        \cline{2-6}
        ~ & Mi  & 99.4 & 97.3 & 3.0 & 2.7 \\
        \cline{2-6}
        ~ & \cellcolor{lightgray} AVG & \cellcolor{lightgray}98.1 & \cellcolor{lightgray}93.8 & \cellcolor{lightgray}7.1 & \cellcolor{lightgray}5.4 \\ 
        \hline
        \hline
    \multirow{5}*{\textbf{Fusion}} & Note & 99.9 & 96.1 & 5.8 & 0.2 \\ 
	\cline{2-6}
	~ & S9  & 99.9 & 98.1 & 1.3 & 0.1 \\
	\cline{2-6}
        ~ & S21 & 99.9 & 99.7 & 0.3 & 0.3 \\
        \cline{2-6}
        ~ & Mi  & 99.9 & 99.1 & 0.6 & 0.1 \\
        \cline{2-6}
        ~ & \cellcolor{lightgray}AVG & \cellcolor{lightgray}\textbf{99.9} & \cellcolor{lightgray}\textbf{98.3} & \cellcolor{lightgray}\textbf{2.0} & \cellcolor{lightgray}\textbf{0.2} \\ 
        \hline
\end{tabular}}
\end{table}

\begin{table}
  \caption{Cross-device face liveness detection results.}
  \vspace{-0.3cm}
  \label{cross_device}
  \centering
  \renewcommand\arraystretch{1.15}
  \scalebox{0.90}{\begin{tabular}{c|c|c|c|c|c}
    \hline
    Modality & Device & AUC(\%)$\uparrow$ & ACC(\%)$\uparrow$ & HTER(\%)$\downarrow$ & EER(\%)$\downarrow$\\
    \hline
    \hline
    \multirow{5}*{\textbf{Vision}}& Note & 99.9 & 98.0 & 1.4 & 0.2 \\
	\cline{2-6}
	~ & S9  & 99.9 & 99.9 & 0.1 & 0.2 \\
	\cline{2-6}
        ~ & S21 & 99.9 & 91.3 & 13.0 & 1.3 \\
        \cline{2-6}
        ~ & Mi & 88.0 & 75.7 & 36.3 & 20.0 \\
        \cline{2-6}
        ~ & \cellcolor{lightgray} AVG & \cellcolor{lightgray}96.9 & \cellcolor{lightgray}91.2 & \cellcolor{lightgray}12.7 & \cellcolor{lightgray}5.4 \\
    \hline
    \hline
    \multirow{5}*{\textbf{Acoustic}} & Note & 74.3 & 58.3 & 36.1 & 30.9 \\
	\cline{2-6}
	~ & S9  & 95.3 & 88.6 & 13.3 & 11.4 \\
	\cline{2-6}
        ~ & S21 & 74.0 & 58.4 & 33.2 & 32.8 \\
        \cline{2-6}
        ~ & Mi  & 96.6 & 70.9 & 43.5 & 9.3 \\
        \cline{2-6}
        ~ & \cellcolor{lightgray}AVG & \cellcolor{lightgray}85.0 & \cellcolor{lightgray}69.0 & \cellcolor{lightgray}31.5 & \cellcolor{lightgray}21.1 \\
        \hline
        \hline
    \multirow{5}*{\textbf{Fusion}} & Note & 99.9 & 98.0 & 1.4 & 0.2 \\
	\cline{2-6}
	~ & S9  & 99.9 & 99.9 & 0.1 & 0.2 \\
	\cline{2-6}
        ~ & S21 & 99.9 & 92.2 & 11.5 & 0.7 \\
        \cline{2-6}
        ~ & Mi & 98.9 & 80.9 & 28.5 & 4.5 \\
        \cline{2-6}
        ~ & \cellcolor{lightgray}AVG & \cellcolor{lightgray}\textbf{99.7} & \cellcolor{lightgray}\textbf{92.8} & \cellcolor{lightgray}\textbf{10.3} & \cellcolor{lightgray}\textbf{1.4} \\
        \hline
\end{tabular}}
\end{table}

\subsubsection{Generalization to unseen variables}
In this section, we examine the robustness of our M$^3$FAS system towards unseen environmental variables. As discussed in Sec. 3.2, the Echoface-Spoof database incorporates variables such as distance, ambient noise, and head pose to accommodate uncontrollable environments when deployed in practical scenarios. In this experiment, we keep the training and testing data from the same device but from different subjects. For each variable ($e.g.$, distance, ambient noise, and head pose), we set three different values (distance: 25cm, 35cm, 45cm; ambient noise: 40dB, 60dB, 70dB; head pose: -10degree, 0degree, 10degree) during the data acquisition. The model is alternately trained on two values and directly tested on the other one, and the average results over three trials are reported in Table~\ref{eval_environment}. Again, we give the detection performance (AUC and HTER) of vision, acoustic, and fusion heads in this table, from which we  observe that the detection performance of the fusion head is superior to the vision and acoustic branch from both perspectives of the average results and the specific device detection performance. This indicates that the M$^3$FAS system successfully fuses the informative RGB and acoustic spoofing cues, even under various challenging cross-environment settings. On the other hand, the proposed model can achieve outstanding 99.1\%, 99.6\%, and 98.7\% AUC scores when crossing towards the unforeseen distance, ambient noise, and head pose, demonstrating the extraordinary generalization capability of the devised system. 

\begin{table}
  \caption{Cross-environment variable evaluation results.}
  \vspace{-0.3cm}
  \label{eval_environment}
  \centering
  \renewcommand\arraystretch{1.15}
  \scalebox{0.73}{\begin{tabular}{c|c|c|c|c|c|c}
    \hline
     \textbf{Vision} & \multicolumn{2}{c|}{Distance} & \multicolumn{2}{c|}{Ambient noise} & \multicolumn{2}{c}{Head pose} \\
    \hline
     \textbf{Device} & AUC(\%)$\uparrow$ & HTER(\%)$\downarrow$ & AUC(\%)$\uparrow$ & HTER(\%)$\downarrow$ & AUC(\%)$\uparrow$ & HTER(\%)$\downarrow$ \\
    \hline
    Note & 94.8 & 12.2 & 96.6 & 12.9 & 94.3 & 12.3\\
    \hline
    S9 & 97.8 & 9.4 & 99.6 & 4.4 & 99.9 & 3.9\\
    \hline
    S21 & 99.1 & 2.4 & 99.9 & 4.9 & 98.5 & 7.9\\
    \hline
    Mi & 99.9 & 4.3 & 99.9 & 12.0 & 98.2 & 11.5\\
    \hline
    \cellcolor{lightgray}AVG & \cellcolor{lightgray}97.9 & \cellcolor{lightgray}7.1 & \cellcolor{lightgray}99.0 & \cellcolor{lightgray}8.6 & \cellcolor{lightgray}97.7 & \cellcolor{lightgray}8.9\\
    \hline
    \hline
    \textbf{Acoustic} & \multicolumn{2}{c|}{Distance} & \multicolumn{2}{c|}{Ambient noise} & \multicolumn{2}{c}{Head pose} \\
    \hline
    \textbf{Device} & AUC(\%)$\uparrow$ & HTER(\%)$\downarrow$ & AUC(\%)$\uparrow$ & HTER(\%)$\downarrow$ & AUC(\%)$\uparrow$ & HTER(\%)$\downarrow$ \\
    \hline
    Note & 93.5 & 12.2 & 99.1 & 4.0 & 96.4 & 16.8\\
    \hline
    S9 & 99.3 & 3.5 & 99.9 & 1.0 & 98.3 & 13.6\\
    \hline
    S21 & 96.2 & 11.2 & 91.5 & 13.9 & 95.8 & 9.6\\
    \hline
    Mi & 97.2 & 6.6 & 99.8 & 1.5 & 98.7  & 7.6\\
    \hline
    \cellcolor{lightgray}AVG & \cellcolor{lightgray}96.6 & \cellcolor{lightgray}8.3 & \cellcolor{lightgray}97.6 & \cellcolor{lightgray}5.1 & \cellcolor{lightgray}97.3 & \cellcolor{lightgray}11.9\\
    \hline
    \hline
    \textbf{Fusion} & \multicolumn{2}{c|}{Distance} & \multicolumn{2}{c|}{Ambient noise} & \multicolumn{2}{c}{Head pose} \\
    \hline
    \textbf{Device} & AUC(\%)$\uparrow$ & HTER(\%)$\downarrow$ & AUC(\%)$\uparrow$ & HTER(\%)$\downarrow$ & AUC(\%)$\uparrow$ & HTER(\%)$\downarrow$ \\
    \hline
    Note & 97.8 & 11.9 & 98.6 & 10.7 & 96.9 & 8.7\\
    \hline
    S9 & 99.3 & 2.3 & 99.9 & 2.2 & 99.9 & 5.4\\
    \hline
    S21 & 99.2 & 7.4 & 99.9 & 3.6 & 99.4 & 6.1\\
    \hline
    Mi & 99.9 & 2.5 & 99.9 & 0.4 & 98.7 & 3.4\\
    \hline
    \cellcolor{lightgray}AVG & \cellcolor{lightgray}\textbf{99.1} & \cellcolor{lightgray}\textbf{6.0} & \cellcolor{lightgray}\textbf{99.6} & \cellcolor{lightgray}\textbf{4.2} & \cellcolor{lightgray}\textbf{98.7} & \cellcolor{lightgray}\textbf{5.9}\\
    \hline
\end{tabular}}
\end{table}

\subsubsection{\textcolor{black}{Robustness} to unseen image distortions}
\begin{figure*}[ht]
\centering
\includegraphics[scale=0.38]{ 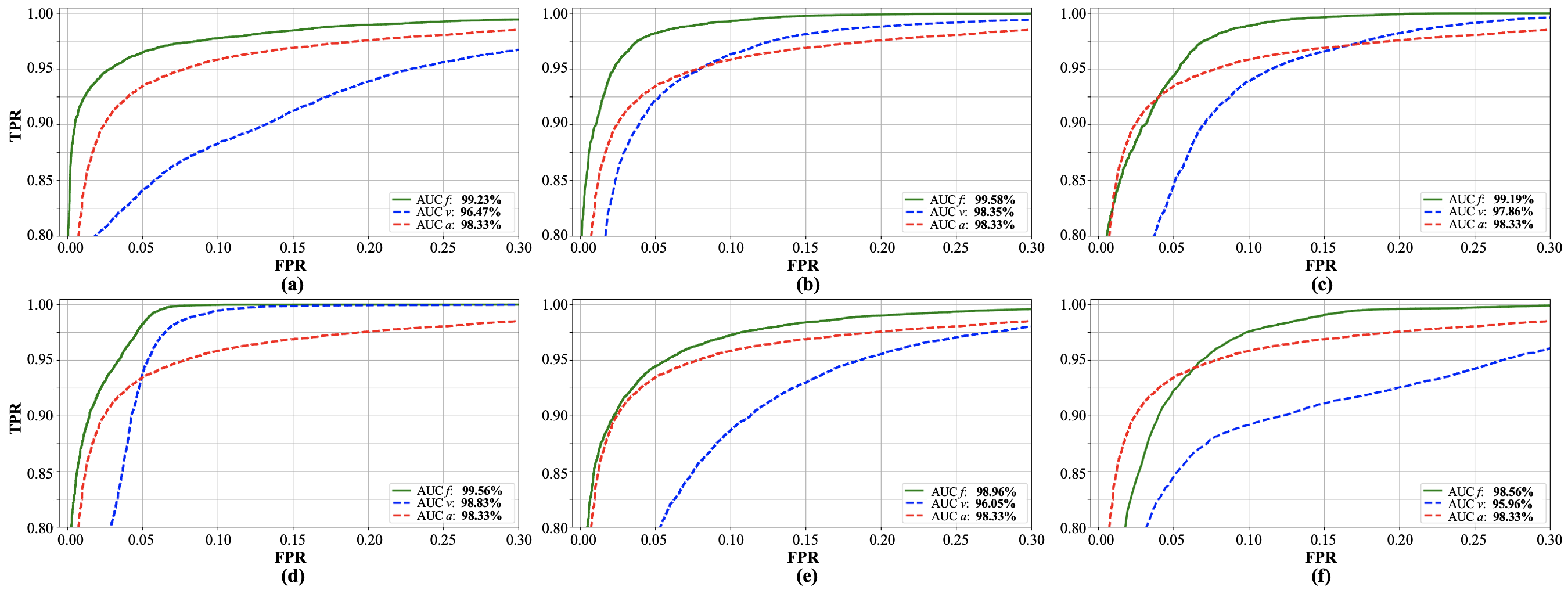}
\vspace{-0.4cm}
\caption{Face liveness detection performance (AUC) to various unseen distortions. (a). Gaussian blur; (b). Image color quantization with dither; (c). JPEG2000 compression; (d). JPEG compression; (e). Pink noise; (f). White noise. The solid green line indicates the fusion result. The vision and acoustic dash lines correspond to the results of the vision and acoustic branches.}
\label{distortion_auc}
\end{figure*}

\begin{figure}[ht]
\centering
\includegraphics[scale=0.35]{ 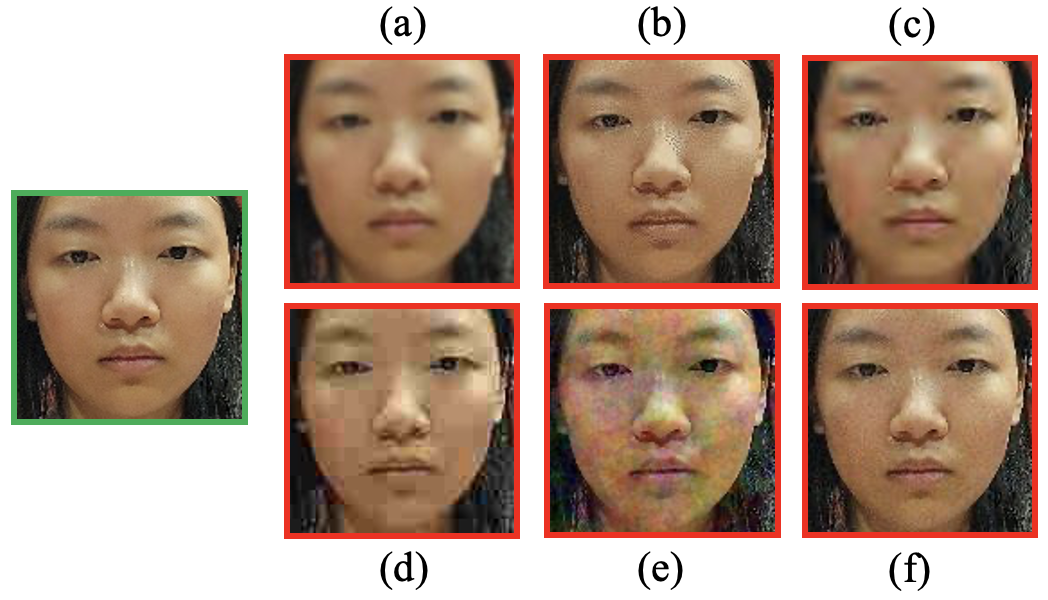}
\vspace{-0.3cm}
\caption{Pristine and distorted face examples. The left pristine face image is boxed in green, and the distorted face images are boxed in red color: (a). Gaussian blur; (b). Image color quantization with dither; (c). JPEG2000 compression; (d). JPEG compression; (e). Pink noise; (f). White noise.}
\label{distortion}
\end{figure}

For deployed face spoofing detectors, robustness to various face image distortions is critical, as hardware degradation and intrinsic noise are inevitable in real-world applications. Shown in Fig.~\ref{distortion} are the incorporated six common image corruption types. The left face highlighted in the green box corresponds to the pristine face picture. Moreover, the face pictures boxed in red represent six common perturbations, Gaussian blur, image color quantization with dither, JPEG2000 compression, JPEG compression, pink noise, and white noise. The model is trained on pristine faces and directly tested on corrupted versions. Compared with the AUC score on pristine face images (99.9\%), Fig.~\ref{distortion_auc} shows that the AUC scores on distorted face images are consistently lower. Besides, the detection performances of the fusion heads are superior to the vision and acoustic branches among all image distortion types, demonstrating the effectiveness and robustness of the proposed multi-modality learning scheme. 



\begin{table}
  \caption{Average computational cost and resource consumption.}
  \vspace{-0.3cm}
  \label{consumption}
  \centering
  \renewcommand\arraystretch{1.15}
  \scalebox{0.9}{\begin{tabular}{c|c|c|c}
    \hline
     Modality & Memory ($MB$) & CPU ($ms$) & Delay ($ms$)\\
    \hline
    Acoustic & 46  & 8 & 48 \\
    \hline
    Vision & 110 & 36 & 83 \\
    \hline
    M$^{3}$FAS & 144 & 73 & 133\\ 
    \hline
\end{tabular}}
\end{table}

\subsubsection{Discussion} 
In this subsection, we examine the robustness of the proposed M$^3$FAS system. The auditory modality captures geometric surface features from input query, thus, can be regarded as a complementary modality to visual inputs for boosting the model's generalization capability. Extensive experiment results demonstrate that the devised system is resilient to various uncontrollable variables, including device, distance, noise, head pose, and image distortions. 

\begin{table}
  \caption{Cross-subject PAD results with different visual branches.}
  \vspace{-0.4cm}
  \label{intra_device_rgb}
  \centering
  \renewcommand\arraystretch{1.15}
  \scalebox{0.83}{\begin{tabular}{c|c|c|c|c|c}
    \hline
    Modality & RGB branch & AUC(\%)$\uparrow$ & ACC(\%)$\uparrow$ & HTER(\%)$\downarrow$ & EER(\%)$\downarrow$\\
    \hline
    \hline
    \multirow{6}*{\textbf{Vision}} & ResNet18 & 99.8 & 96.7 & 3.7 & 1.8\\ 	
	\cline{2-6}
	~ & DenseNet161 & 99.9 & 98.3 & 1.3 & 0.7\\ 
	\cline{2-6} 
        ~ & CDCN & 99.1 & 94.8 & 4.7 & 4.2 \\	
        \cline{2-6}
         ~ & ParC-Net-S & 99.8 & 97.9 & 2.8 & 1.0 \\	
        \cline{2-6}
        ~ & MobileNet-v2 & 99.6 & 95.1 & 4.7 & 2.4 \\	
        \cline{2-6}
        ~ & Ours & 99.9 & 97.8 & 2.8 & 0.4 \\  
    \hline
    \hline
    \multirow{6}*{\textbf{Fusion}} & ResNet18  & 99.4 & 97.6 & 3.2 & 2.7\\ 
	\cline{2-6}
	~ & DenseNet161 & 99.9 & 98.8 & 1.6	& 0.1\\
	\cline{2-6} 			
        ~ & CDCN & 99.9 & 97.2 & 2.4 & 1.1\\
        \cline{2-6}
        ~ & ParC-Net-S & 99.9 & 98.5 & 1.8 & 0.5 \\	
        \cline{2-6}
        ~ & MobileNet-v2 & 99.9 & 97.9 & 1.9 & 0.7 \\	
        \cline{2-6}
        ~ & Ours & 99.9 & 98.3 & 2.0 & 0.2\\
        \hline 			
\end{tabular}}
\end{table}


\vspace{-3mm}
\subsection{Flexibility examination of M$^3$FAS} 
\vspace{-1mm}
Although accurate and robust, typical multi-modality models cannot work appropriately in some usage scenarios. For instance, in the M$^3$FAS system, either visual or auditory modality data could be missing due to unexpected hardware breakdown or extreme environmental conditions. Besides, the system can avoid the power-hungry camera in the ultra-low-power mode and solely conduct PAD using the acoustic branch.
The flexibility of our system lies in three aspects: (1). M$^3$FAS can still work properly when one modality is missing ($e.g.$, ultra-low-power mode); (2). M$^3$FAS can respond fast with limited computational resource consumption and thus can be flexibly deployed on commodity mobile devices; (3) M$^3$FAS can flexibly adapt to SOTA RGB-based methods and improve the existing methods' PAD performance.  

\subsubsection{Computational resource consumption}
We deploy our M$^3$FAS system on a \texttt{Samsung s21} and report memory ($MB$), CPU ($ms$), and Delay ($ms$) of the acoustic branch, vision branch, and M$^3$FAS in Table~\ref{consumption}. 
We evaluate the main memory cost during the recording, processing, and classification. The memory footprint has an average $\sim$46$MB$, $\sim$110$MB$, and $\sim$144$MB$ for acoustic, vision, and fusion. The time for the CPU to complete the machine learning inferences is low in acoustic (8$ms$), vision (36$ms$), and M$^3$FAS (73$ms$) modes.
The response delays represent the time for the system to produce the spoofing results after receiving the raw image/signal data, which are 48$ms$, 83$ms$, and 133$ms$, respectively. \textcolor{black}{The usage of acoustic modality brings extra computational costs and resource consumption (34 MB memory, 37 ms inference time, and 50 ms response delay on the \texttt{Samsung s21}). However, we posit that this is an acceptable trade-off, as it has a negligible impact on the user experience during the authentication process. Moreover, with the rapid advancements in built-in hardware, approximately 30 MB of additional memory usage will pose no significant challenge for commercial-off-the-shelf (COTS) mobile devices. In addition, the inference time and response delay can also be considerably reduced on emerging high-end mobile devices.} Naturally, the M$^{3}$FAS system has the highest resource consumption, but it may also have improved performance compared to using a single modality due to the integration of multiple sources of information.
M$^3$FAS system is feasible to be implemented into mobile devices. 

\begin{table}
  \caption{Computational costs and model sizes of different visual branches.}
  \vspace{-3mm}
  \label{parameters}
  \centering
  \renewcommand\arraystretch{1.15}
  \scalebox{1.0}{\begin{tabular}{c|c|c}
    \hline
      & MACs &  \#param. \\
    \hline
    ResNet18  & 1.82 G & 11.18 M \\
    \hline
    DenseNet161 & 10.24 G & 26.48 M \\
    \hline 	
    CDCN & 35.46 G & 3.21 M \\
    \hline 	
    ParC-Net-S & 3.50 G & 5.00 M \\
    \hline 
    MobileNet-v2 & 0.30 G & 3.47 M \\
    \hline 	
    \hline
    Ours & 1.17 G & 2.58 M \\ 	
    \hline
\end{tabular}}
\end{table}

\vspace{-4mm}
\textcolor{black}{\subsubsection{Effectiveness of the designed framework with existing RGB methods}}
\textcolor{black}{M$^3$FAS can flexibly adapt to existing RGB-based FAS methods. In this subsection, we adopt five representative network architectures as alternatives for the vision branch: ResNet18~\cite{he2016deep}, DenseNet161~\cite{huang2017densely}, CDCN~\cite{yu2020searching}, ParC-Net-S \cite{zhang2022parc}, and MobileNet-v2 \cite{sandler2018mobilenetv2}. We report the cross-subject evaluation results in Table 10, where ``Vision" indicates the exclusive usage of RGB modality data for FAS. We integrate existing RGB architectures with our designed acoustic branch for the fusion modality evaluation while maintaining flexible modal learning and hierarchical cross-modal learning schemes to ensure a fair comparison.}

\textcolor{black}{From the vision results, it can be readily observed that our visual branch achieves the best AUC and EER performance and the second-best ACC and HTER scores, demonstrating its effectiveness for FAS. By incorporating the acoustic branch with the RGB branches, all the listed methods consistently demonstrate certain performance enhancement, underscoring the effectiveness of our designed framework. While DenseNet161 achieves the best FAS performance among all listed approaches, our method still maintains competitive AUC and ACC scores compared to DenseNet161. This observation is reasonable since DenseNet161 is a considerably larger model that demands more computational resources for training and testing.} \textcolor{black}{We also incorporate typical lightweight networks, ParC-Net-S \cite{zhang2022parc} and MobileNet-v2 \cite{sandler2018mobilenetv2}, to assess the effectiveness of the proposed learning scheme on mobile networks. The results from both the vision and fusion experiments further demonstrate that the proposed audio-visual FAS method can effectively enhance the FAS performance of mobile networks.}

\textcolor{black}{Additionally, we have measured the computational costs and model sizes of each RGB branch in Table 11. Notably, our vision branch exhibits significantly reduced computational demands and a more compact model size than other RGB-based FAS models. \textcolor{black}{Table 11 indicates that despite the inclusion of typical mobile networks like ParC-Net-S and MobileNet-v2, our vision branch maintains a more parameter-efficient network.}  Therefore, our method tends to be a better choice than RGB-based models, particularly for mobile applications.} 

\subsubsection{Discussions and Justifications}
\textcolor{black}{The designed M$^{3}$FAS is a user-friendly, secure, privacy-preserving, and robust FAS system. Our system emits a customized signal through the speaker, ensuring that it does not access other sound files stored on the user's mobile phone. This approach safeguards sensitive sound information from being leaked. Additionally, our method does not require additional user interaction, such as speaking, thereby enhancing the overall user experience. Permissions granted to users adhere to the standard permission-granting process of the Android system. Modifying application permissions involves only a few simple steps, ensuring minimal inconvenience to users.}

\textcolor{black}{Compared to existing RGB-based methods, our M$^{3}$FAS system exhibits a more generalized and robust FAS method due to its incorporation of both RGB texture information and acoustic geometric information. Besides, the M$^{3}$FAS system is more user-friendly compared to other speaking authentication methods because it does not require users to speak or perform specific actions. Moreover, it provides a more flexible FAS system, which can still work properly when the users do not grant the access permissions to specific sensors.} 

\textcolor{black}{Our system will not cause security issues or privacy leakage. Because users can choose to grant one-time permission, which only allows the application to access the requested sensors during the current runtime.} 

\textcolor{black}{The M$^{3}$FAS system demonstrates outstanding robustness in various human daily activities, even amidst various noise levels. This resilience is ensured by the utilization of a high-pass filter and a recording frequency range of smartphones, jointly guaranteeing a high Signal-to-Noise Ratio (SNR) in the recorded acoustic signals. Experimental results provided in Table 8 show the robustness of our proposed method.}

\section{Conclusion and future work}
This paper presented a user-friendly and secure face anti-spoofing system, M$^3$FAS, for mobile devices. M$^3$FAS effectively combines RGB texture and acoustic geometric spoofing clues, thereby achieving accurate and robust PAD performances. We designed a two-modality neural network, which, for the first time to the best of our knowledge, smartly fuses vision and acoustic features for PAD. The proposed hierarchical cross-attention modules perform  cross-modal feature learning from three levels, contributing to better detection performances. Finally, we found that our multi-head learning strategy could mitigate the overfitting problem and enable a more flexible PAD. Extensive experiments demonstrated that the proposed M$^3$FAS system could perform secure, robust, and flexible PAD.


While the proposed FAS system is robust under a wide variety of experimental settings, the recorded high-frequency acoustic signals differ dramatically among different devices, resulting in a limited performance in cross-device evaluations for single auditory modality. This domain gap is mainly caused by the hardware ($i.e.$, speaker and microphone) discrepancies among different mobile devices. In future work, applying effective signal processing and domain generalization algorithms to extract device-independent spoofing features opens an important research path forward. Pruning less important layers and neurons to achieve model compression is also essential for mobile deployment. Moreover, adapting our FAS system to 3D face presentation attacks is worth considering. We envision that M$^3$FAS should still be effective since the auditory modality can reflect informative reflectance features of bonafide and 3D attacks but that requires further investigation.



\ifCLASSOPTIONcaptionsoff
  \newpage
\fi

\bibliographystyle{IEEEtran}
\bibliography{main}

\end{document}